\documentclass[preprint,12pt]{elsarticle}
\usepackage{amssymb}
\usepackage{amsmath} 
\usepackage{textcomp}
\usepackage{caption}
\usepackage{subcaption}
\usepackage{siunitx}
\journal{Int. J. Pharm.}

\begin{document}

\begin{frontmatter}



\title{A computational fluid dynamics model for the simulation of flash-boiling flow inside pressurized metered dose inhalers}


\author[inst1]{Riccardo Rossi\corref{cor1}}
\ead{rrossi@red-fluid.com}
\author[inst2]{Ciro Cottini}
\author[inst2,inst3]{Andrea Benassi}
\cortext[cor1]{Corresponding author}
\affiliation[inst1]{organization={RED Fluid Dynamics},city={Cagliari},country={Italy}}
\affiliation[inst2]{organization={Chiesi Farmaceutici S.p.A.},city={Parma},country={Italy}}
\affiliation[inst3]{organization={International School for Advanced Studies (SISSA)},city={Trieste},country={Italy}}

\begin{abstract}
{In this work we present, for the first time, a computational fluid dynamics tool for the simulation of the metered discharge in a pressurized metered dose inhaler. The model, based on open-source software, adopts the Volume-Of-Fluid method for the representation of the multiphase flow inside the device and a cavitation model to explicitly account for the onset of flash-boiling upon actuation. Experimental visualizations of the flow inside the device and measurements of the mixture density and liquid and vapor flow rates at the nozzle orifice are employed to validate the model and assess the sensitivity of numerical results to modeling parameters. The results obtained for a standard device geometry show that the model is able to quantitatively predict several aspects of the dynamics and thermodynamics of the metered discharge. We conclude by showing how, by allowing to reproduce and understand the fluid dynamics upstream of the atomizing nozzle, our computational tool enables systematic design and optimization of the actuator geometry.} 
\end{abstract}

\begin{graphicalabstract}
\includegraphics[width=1.0\textwidth,trim={0 0 0 0},clip]{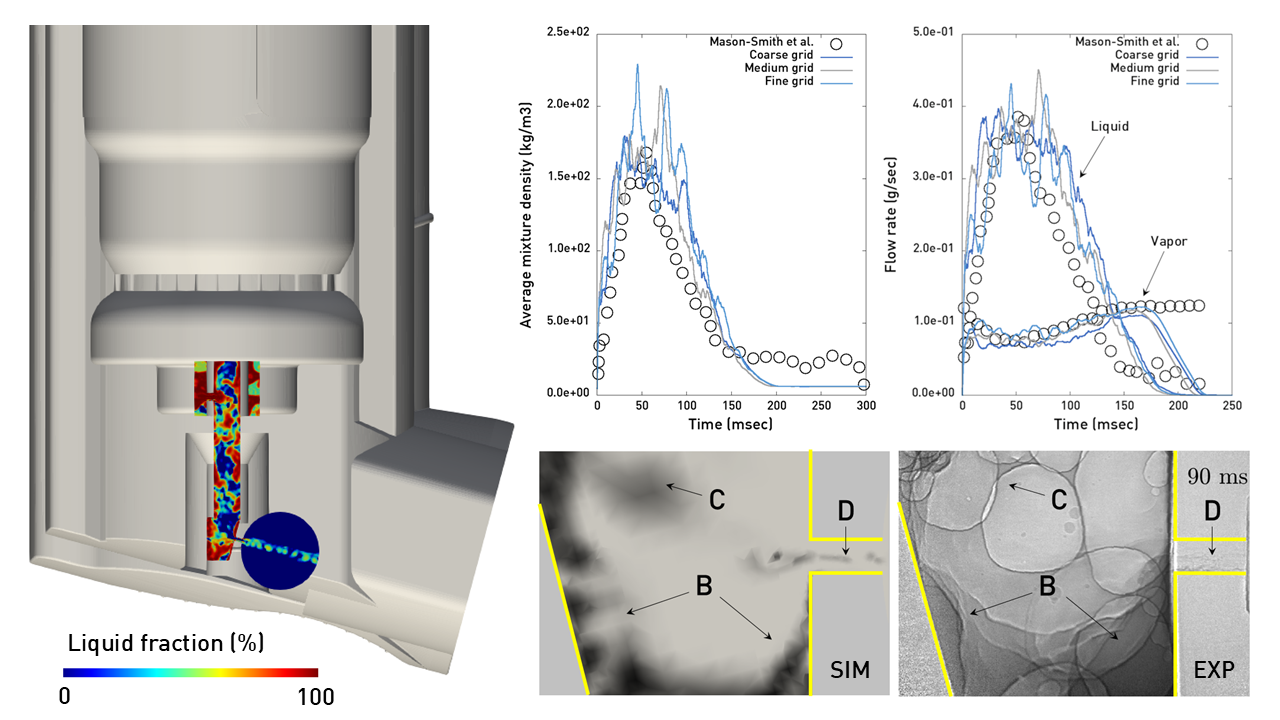}
\end{graphicalabstract}

\begin{highlights}
\item {The CFD model successfully simulates the metered discharge in a pMDI}
\item {Good agreement with experiments is found for mixture profiles at nozzle orifice}
\item Equilibrium cavitation models are able to represent the flash-boiling flow
\item Turbulent fluctuations do not impact significantly the metered discharge
\item Effects of geometrical modifications on metered discharge are captured by the CFD model
\end{highlights}


\end{frontmatter}


\section{Introduction}\label{introduction}
The wide adoption of pressurized Metered Dose Inhalers (pMDI) for the administration of orally inhaled drug products stems from their low cost, reasonable delivery efficiency and versatility. In fact, Dry Powder Inhalers (DPIs) have much higher costs, while nebulizers require long administration times, need to be connected to a plug or a battery and usually lead to large drug waste. Being the aerosol generated through a propellant, the delivery performance of pMDIs are less sensitive to the patient inspiratory effort than DPIs. However, pMDIs also come with limitations: first, they require the patient to coordinate its inspiratory act with the triggering of the inhaler and secondly, they make use of propellants and solvents, such as ethanol, and they are thus not adequate to deliver bio-molecules~\cite{Finlay2019,DeBoer2021a,DeBoer2021b,Smyth2006}.

Many factors influence the pMDIs inhalation performance. The choice of the propellant determines the maximum pressure achieved during actuation upstream of the atomizing nozzle, impacting the initial droplet size~\cite{Zhu2015}, whereas the concentration and physico-chemical properties of the solvent determine the droplet evaporation time and thus the progressive reduction of the Mass Median Aerodynamic Diameter (MMAD) of the airborne aerosol~\cite{LeghLand2021,Sheth2017,Stein2006}. Propellant and solvent choice, as well as the solvent concentration in the mixture, affect plume velocity and plume angle, determining how much aerosol is held back in the mouth-throat tract~\cite{Yousefi2017,Kakade2007}. Design and optimization of pMDI products could definitely become faster and cheaper if these aerosol characteristics could be predicted, using mathematical models and simulations, solely based on the drug product composition and its physico-chemical properties. However, aerosol droplet formation in an atomizing nozzle is a complex physical phenomenon to study, different types of mechanical instabilities are at play depending on the dimensionless Weber number, i.e. the ratio between the droplet inertial force and its surface tension force. Both experimental imaging and simulation of droplet formation are very demanding tasks, this is why many empirical correlations have been developed to predict the MMAD of the generated droplets based on the nozzle geometry and the fluid mixture properties. Most used, among them, is the Clark correlation~\cite{Clark1991}:        

\begin{equation}\label{eqn:ClarkCorrelation}
    MMAD=\frac{C}{\chi^{0.56}\left[\left(p_{in}-p_{amb}\right)/p_{amb}\right]^{0.46}}
\end{equation}

with $p_{in}$ being the pressure upstream of the nozzle, i.e. in the expansion chamber, $p_{amb}$ the ambient pressure downstream of the nozzle, $\chi$ is the vapor quality factor (the mass of vapor divided by the total mass of the mixture in the unit volume) and $C$ is a fitting constant equal to $1.82$ in case of a metered discharge. The equation contains no explicit nozzle geometric parameter; however, the pressure drop between the expansion chamber and the ambient pressure depends implicitly on the nozzle diameter and takes different values whether the flow at the nozzle is subsonic, critical (choked) or supersonic. Other simple equations have been developed to estimate the droplet initial velocity and the aerosol production rate~\cite{Zhu2015,Clark1991,Chizari2023}. As for the Clark correlation~(\ref{eqn:ClarkCorrelation}), they all depend on dynamics and thermodynamics properties of the mixture upstream of the nozzle. Therefore, the problem of predicting pMDIs delivery performance requires the capability to calculate the aerosol physical characteristics and, in turn, the prediction of mixture properties inside the pMDI during administration. A first attempt in this sense was made in 1975 by Fletcher~\cite{fletcher1975} through a mathematical model of the metered discharge, subsequently extended and improved by many authors~\cite{Clark1991,Dunbar1997,Ju2010,Gavtash2017,Gavtash2018}. System models similar to the one of Fletcher, where a zero or one-dimensional representation of the pMDI is introduced, allow to predict the temporal evolution of the mixture properties by solving the integral balance equations for the mixture in the main device components. Several assumptions need to be made in the framework of these models, such as the type of flow regime and the value of discharge coefficients through the orifices. Moreover, system models are unable to study the effect of design changes on the pMDI performance and operation.

Most of the experimental effort is devoted to the characterization of the generated aerosol plume in terms of droplets size and speed, using laser light scattering techniques such as phase Doppler anemometry~\cite{Alatrash2016} and particle image velocimetry~\cite{Crosland2009}, or in terms of plume geometry and temperature~\cite{Jahed2024,Kotak2020,MoragaEspinoza2018}. Little is known about the flash-boiling of propellant and the fluid dynamics upstream of the atomizing nozzle, i.e. inside the actuator. In this sense a seminal paper is the one by Mason-Smith et al.~\cite{MasonSmith2017}, where X-ray phase contrast imaging and quantitative radiography have been used to investigate the behavior of the solvent-propellant mixture during actuation, while the liquid and mass flow rate at the nozzle outlet have been quantified optically. The same imaging technique has been adopted more recently by McKiernan~\cite{McKiernan2019} to further study the structure and dynamics of the flashing flow in the pMDI, but without providing quantitative information on the mixture during the metered discharge. Temperature profiles during the shot were measured by Gavtash et al.~\cite{Gavtash2019} and Myatt et al.~\cite{Myatt2022} in the sump and nozzle-orifice, respectively, whereas Myatt et al.~\cite{Myatt2015a,Myatt2015b} and Gavtash~\cite{Gavtash2016} reported measurements for the near-orifice mixture velocity downstream of the nozzle orifice.

In the context of pMDI design, Computational Fluid dynamic (CFD) simulations are usually applied to study the behavior of the aerosol plume downstream of the device~\cite{Xi2022,Oliveira2014} and the consequent aerosol deposition in patient mouth-throat tract and bronchial tree~ \cite{Dastoorian2022,Ahookhosh2021,Yousefi2017,Walenga2013}. Here, for the first time, we demonstrate the possibility of using a multiphase CFD model to simulate the metered discharge upon actuation, capturing the propellant dynamics in time and in a realistic three-dimensional geometry for the actuator, from the metering chamber to the nozzle orifice. Details on the simulated geometry and pMDI operating conditions are illustrated in Section~\ref{section:deviceGeometry}. The computational model is presented in Section~\ref{section:computationalModel} along with the physical and numerical assumptions implied in its development. The Results section opens with a qualitative description of the metered discharge and its comparison with available experimental imaging data, and continues with the model calibration step, a grid-sensitivity study and by analyzing the effect of including temperature dependence of the physical properties of the mixture as well as the latent heat due to phase-change. A final section illustrates how the simulation tool can be used to predict the effects on the metered discharge from modifications in the actuator geometry and volume. The Conclusions section summarizes the outcome of the present study, the open points and the need of new measurements to challenge and extend the predictive capabilities of the model. 

\section{Device geometry and operating conditions}\label{section:deviceGeometry}
\begin{figure}
\centering
\includegraphics[width=1.0\textwidth,trim={0 0 0 0},clip]{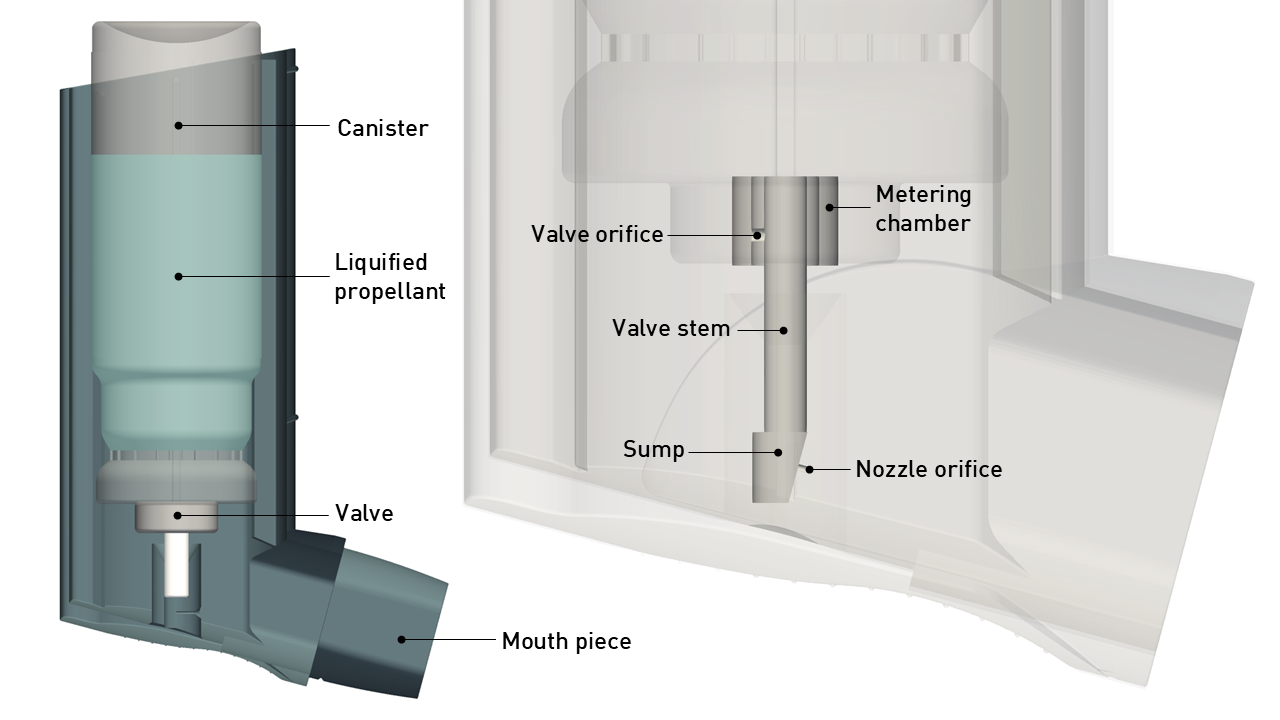}
\caption{Overview of the pressurized metered dose inhaler (left) and main device components involved in the metered discharge (right).}
\label{fig:introduction:deviceGeometry:deviceGeometry}
\end{figure}

In a pMDI device, shown in Fig.~\ref{fig:introduction:deviceGeometry:deviceGeometry}, the Active Pharmaceutical Ingredient (API) is solubilized or suspended in a liquid mixture composed by a propellant, usually a hydrofluoroalkane (HFA) and by a solvent, typically ethanol. The mixture is contained inside a metallic canister in equilibrium with its vapor phase, i.e. at its vapor pressure, typically between 3 and 5 bar. A valve is crimped on the canister with a metering chamber whose aim is to collect the dose volume, usually few tens of \SI{}{\micro\liter}. Upon actuation, the dose is isolated from the canister and exposed to ambient pressure, thereby inducing the so-called flash-boiling phenomenon, an extremely rapid and violent phase-change process during which the liquid and vapor mixture is initially flushed from the metering chamber into the valve stem through the valve orifice and subsequently collected into the sump region. From this region, the mixture enters the nozzle orifice undergoing atomization, which ultimately generates the aerosol delivered to the patient through the mouthpiece.

In the present work, the metered discharge of a \SI{50}{\micro\liter} dose of pure HFA-134a propellant is investigated {by numerical simulations. The size of the metered valve and the device internal geometry corresponds to the one adopted in the experimental work} by Mason-Smith et al.~\cite{MasonSmith2017}, which will represent the main reference experiment for the validation of the numerical model presented in this study. The valve and nozzle orifice diameters are \SI{0.6}{\milli\meter} and \SI{0.3}{\milli\meter}, respectively, and the valve stem is \SI{12}{\milli\meter} and \SI{2}{\milli\meter} in length and diameter, respectively. The propellant is stored at 5.9 bar, representing the nominal saturation pressure of the HFA134a at room temperature of \SI{20}{\celsius}. From the experiments, a typical duration of the metered discharge between \SI{250}{\ms} and \SI{300}{\ms} is observed, which will be replicated in the simulations.

\section{Computational model}\label{section:computationalModel}
The computational model of the pMDI consists of a numerical representation in both space and time of the flow inside the device. Compared to most advanced system models~\cite{Gavtash2018}, where only the temporal evolution of the quantities of interest can be predicted, the model is also capable of providing a spatial description of such quantities during the metered discharge. This allows to extend the predictive capabilities of simulation tools to the effect of modifications to the pMDI internal geometry, yet to be explored, on device operation and performance.

The model is developed using the open-source software OpenFOAM$^\text{\textregistered}$~\cite{Weller1998}, based on the unstructured finite-volume method, where the flow governing equations are solved on a computational grid consisting of a finite number of cells representing the volume inside the metered-valve, valve stem and sump regions. In the following sections, the governing equations of the multiphase flow model and of the model adopted for the simulation of the flash-boiling phenomenon are presented, along with the computational grid and numerical methods employed to solve the equations.

\subsection{Multiphase flow model}\label{subsection:computationalModel:multiphaseModel}
The flow inside a pMDI is represented by a two-phase (liquid and vapor) mixture originating from the flash-boiling occurring inside the metered valve upon device operation. In the present work, a Homogeneous Fluid Model (HFM) is employed to represent the mixture, where mechanical and thermodynamic equilibrium is assumed between the two phases, which thereby share the same pressure, temperature and velocity. Based on these assumptions, a single set of governing equations consisting of the continuity, momentum and energy equations can be solved for the mixture:

\begin{eqnarray}
\frac{\partial\rho}{\partial t}+\frac{\partial\left(\rho U_j\right)}{\partial x_j} & = & S_m\label{eqn:continuityEquation} \\
\frac{\partial\left(\rho U_i\right)}{\partial t}+\frac{\partial\left(\rho U_iU_j\right)}{\partial x_j} & = & -\frac{\partial P}{\partial x_i}+\frac{\partial}{\partial x_j}\left[\left(\mu+\mu_t\right)\left(\frac{\partial U_i}{\partial x_j}+\frac{\partial U_j}{\partial x_i}\right)\right] \nonumber \\
& & + \rho g_i+f_{\sigma}\label{eqn:momentumEquation} \\
\frac{\partial\left(\rho T\right)}{\partial t}+\frac{\partial\left(\rho U_jT\right)}{\partial x_j} & = & -\frac{\partial}{\partial x_j}\left(U_jP\right)+\left(\lambda+\lambda_t\right)\frac{\partial^2 T}{\partial x_j\partial x_j}+S_T\label{eqn:energyEquation}
\end{eqnarray}

where $\rho$, $p$, $U_j$ and $T$ are the density, pressure, velocity and temperature of the mixture, $\mu$ and $\lambda$ are the molecular viscosity and thermal conductivity and $\mu_t$ and $\lambda_t$ are the turbulent viscosity and thermal conductivity associated with the modeling of the turbulent regime of the mixture flow, which will be discussed in Section~\ref{subsection:computationalModel:turbulenceModeling}. The terms $\rho g_i$ and $f_{\sigma}$ in Eq.~(\ref{eqn:momentumEquation}) represent the gravity and surface tension forces, whereas the source terms $S_m$ and $S_T$ on the right-hand side of Eqs.~(\ref{eqn:continuityEquation}) and~(\ref{eqn:energyEquation}) represent the effect of the phase-change due to the flash-boiling of mixture pressure and temperature, which will be introduced in Section~\ref{subsection:computationalModel:flashBoilingModel}.

\begin{figure}
\centering
\includegraphics[width=1.0\textwidth,trim={0 100 0 0},clip]{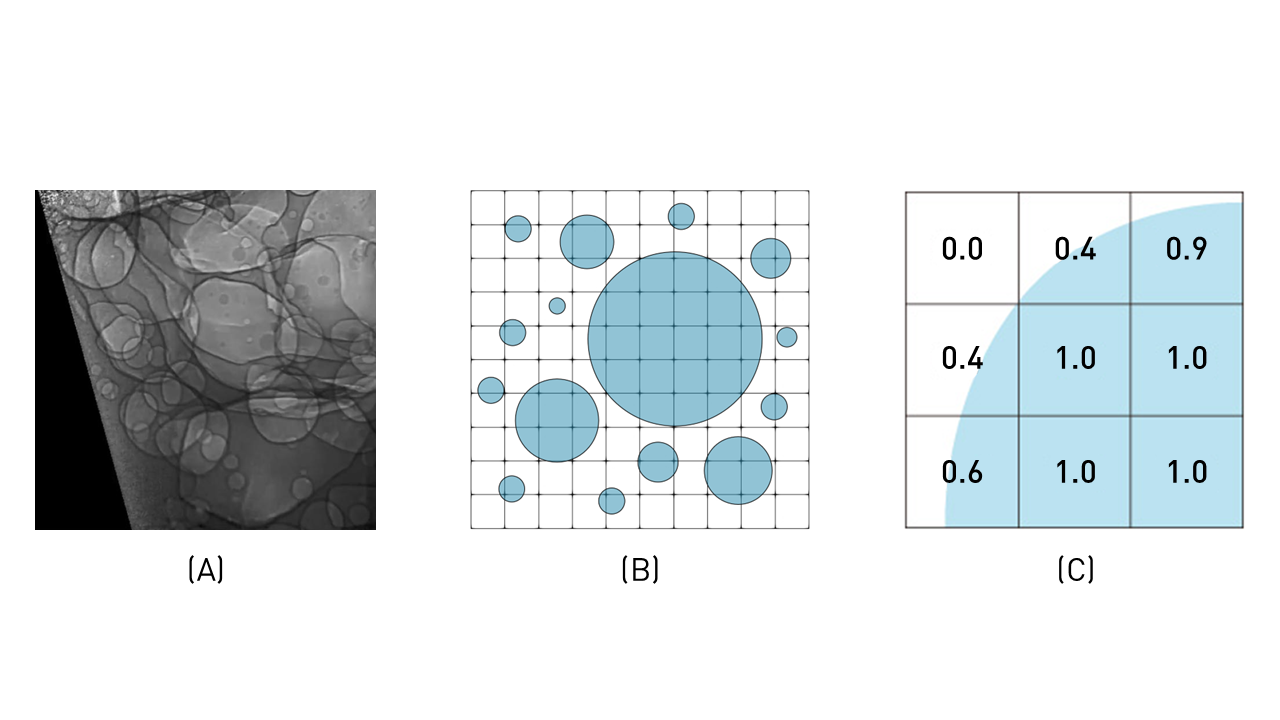}
\caption{Volume-fraction based representation of the multiphase flow inside the pMDI: (a) X-ray visualization of bubbly flow in the nozzle region of the device extracted by the original movie (courtesy of Mason-Smith et al.~\cite{MasonSmith2017}) (b) sketch of bubbly flow on a finite-volume grid, (c) visualization of the liquid volume fraction in a finite-volume grid.}
\label{fig:computationalModel:multiphaseModel:volumeFractionConcept}
\end{figure}

Flow visualizations inside the pMDI performed with X-ray phase contrast imaging~\cite{MasonSmith2017} reveal a dispersed regime in the two-phase flow, where numerous bubbles presenting a broad range of scales are present within the propellant undergoing flashing, as shown in Fig.~\ref{fig:computationalModel:multiphaseModel:volumeFractionConcept}(a). An explicit model of such flow regime, where bubbles with significantly different length-scales must be solved directly, is computationally prohibitive due to the grid resolution needed to capture the liquid-vapor interfaces, as depicted in the cartoon (b) in the same figure. A simplified representation of the two-phase flow based on the Volume-Of-Fluid (VOF) approach~\cite{Nichols1975} is thus adopted here, where the {\em volume-fraction} concept is introduced to estimate the amount of liquid and vapor phase present in each computational cell, as shown in Fig.~\ref{fig:computationalModel:multiphaseModel:volumeFractionConcept}(c).

The volume-fractions of the liquid and vapor phases of the mixture are defined as $\alpha_l=V_l/V$ and $\alpha_v=V_v/V$, respectively, where $V_l$ and $V_v$ denote the volume occupied by the liquid and the vapor phase inside the volume $V$ of the computational cell. In the present model, an additional transport equation is solved to predict the volume fraction of the liquid phase inside the device:

\begin{equation}\label{eqn:volumeFractionEquation}
    \frac{\partial \alpha_l}{\partial t}+\frac{\partial}{\partial x_j}\left(\alpha_lU_j\right)+\frac{\partial}{\partial x_j}\left(\alpha_v\alpha_lU_{rj}\right)=S_{\alpha_l}
\end{equation}

where $\alpha_v$ is the vapor volume-fraction and where the source term $S_{\alpha_l}$, which will also be presented in Section~\ref{subsection:computationalModel:flashBoilingModel}, represents the effect of phase-change associated with the flash-boiling on the liquid volume-fraction. The third term on the left-hand side of Eq.~(\ref{eqn:volumeFractionEquation}) is an artificial compression term used to resolve the interface between the two-phases. Given the dispersed flow regime associated with the pMDI, such term is kept in the present formulation for sake of numerical stability, but the liquid-vapor interface cannot be captured accurately because of the limited grid resolution adopted in the calculations, aimed at reducing the computational effort.

The properties of the mixture are calculated based on the local volume fraction of liquid and vapor. Due to mass-conservation, the volume fractions of the mixture obey to $\alpha_l+\alpha_v=1$. Therefore, only the transport equation for the liquid volume fraction~(\ref{eqn:volumeFractionEquation}) can be solved and the properties of the mixture calculated using the following relations:

\begin{eqnarray}\label{eqn:temperatureDependentProperties}
    \rho=\rho_l\alpha_l+\rho_v\left(1-\alpha_l\right)\nonumber \\
    \mu=\mu_l\alpha_l+\mu_v\left(1-\alpha_l\right)\\
    D=D_l\alpha_l+D_v\left(1-\alpha_l\right)\nonumber
\end{eqnarray}

The properties of the liquid and vapor phase are obtained from the measurements of Tillner-Roth and Baeh~\cite{Tillner-Roth1994}, which characterize the propellant over a wide range of temperatures and pressures. In the calculations, the densities of both vapor and liquid are assumed variables to allow for the simulation of the compressible discharge of the propellant from the metered valve. The vapor density is computed via the ideal-gas equation, whereas a reciprocal polynomial function is employed to compute the liquid density based on the pressure and temperature of the mixture:

\begin{equation}\label{eqn:liquidDensity}
    \frac{1}{\rho_l}=C_0+C_1T+C_2T^2-C_3p-C_4pT    
\end{equation}

where a fit to measured data~\cite{Tillner-Roth1994} is performed to compute the coefficients of Eq.~(\ref{eqn:liquidDensity}). Separate calculations are performed considering the remaining properties of both phases, i.e. the dynamic viscosity, specific heat capacities and thermal conductivity, either constant or temperature-dependent to study the effect on the simulated metered discharge. In the case of constant properties, the values measured at room temperature are employed in the computations whereas look-up tables are implemented in the computational model to obtain the values at run-time in the case of temperature-dependent properties.

\subsection{Turbulence modeling}\label{subsection:computationalModel:turbulenceModeling}
The flow governing equations~(\ref{eqn:continuityEquation})-(\ref{eqn:energyEquation}) are presented in the RANS (Reynolds-Averaged Navier-Stokes) form. In this formulation, the equations are solved for the mean flow variables and the entire spectrum of velocity and temperature fluctuations is modeled via the turbulent viscosity $\mu_t$ and the turbulent thermal conductivity $\lambda_t$, which account for the effect of the turbulent fluctuations on the momentum and energy transport, respectively. Preliminary calculations performed without including the turbulent terms in the governing equations show that the Reynolds number $Re=UD/\nu$ associated with the flow through the expansion chamber and the valve and nozzle orifices, where $\nu=\mu/\rho$ is the kinematic viscosity and $D$ is the diameter of each component, falls in the range of critical values for pipe flows ($Re\sim 2\cdot 10^3$) and therefore a fully turbulent regime should not be expected. Nonetheless, the flash-boiling may induce significant perturbations in the mixture flow across the valve orifice and inside the expansion chamber which, in turn, may lead to the onset of turbulence even at low Reynolds numbers. Furthermore, turbulent fluctuations do not only affect the transport of momentum and energy in flashing-boiling flows, but also have an influence on saturation pressure, which is found increased by the effect of pressure fluctuations~\cite{Singhal2002}:

\begin{equation}\label{eqn:modifiedSaturationPressure}
    P_s=P_v+P_t/2
\end{equation}

where $P_s$ is the corrected saturation pressure, $P_v$ is the nominal vapor pressure of the fluid and $P_t$ the intensity of turbulent pressure fluctuations, which can be estimated as follows:

\begin{equation}\label{eqn:pressureFluctuationsModeling}
    P_t=0.39\rho k
\end{equation}

where $\rho$ is the mixture density and $k$ the turbulent kinetic energy. In the present model, the potential onset of a turbulent regime is simulated using the $k-\omega$ SST model by Menter~\cite{Menter1994}, where two additional transport equations are solved for the the turbulent kinetic energy $k$ and its specific dissipation rate $\omega$ to obtain the turbulent viscosity as $\nu_t=\rho k/\omega$. For the turbulent thermal conductivity, the Reynolds analogy with molecular transport is invoked via the relation $\lambda_t=\mu_tc/Pr_t$, where $c$ is the specific heat capacity of the mixture and $Pr_t$ the turbulent Prandtl number~\cite{Kays1994}, set to 1.0 in the present model.  

\subsection{Flash-boiling model}\label{subsection:computationalModel:flashBoilingModel}
Flash-boiling is the main physical phenomenon driving the discharge of the pMDI as well as the main modeling challenge in the numerical simulation of this device. During flash-boiling, the pressurized propellant stored in the metered chamber undergoes rapid depressurization below the saturation pressure, thus leading to phase-change and to the abrupt formation of the liquid-vapor mixture. The very small time-scales characterizing the flash-boiling and, in the case of the pMDI, the limited volume over which the process takes place, lead to significant departures from mechanical as well as thermal equilibrium between the liquid and vapor phases~\cite{Liao2017}. Nonetheless, several attempts to perform CFD simulations of flash-boiling flows can be found in the literature using models based on the thermal equilibrium assumption for the phase-change, originally developed for cavitating flows. Karathanassis et al.~\cite{Karathanassis2017} compared, among others, the model by Zwart et al.~\cite{Zwart2004}, based on the Rayleigh-Plesset (RP) equation for bubble-growth~\cite{Brennen1995} to the model by Theofanous et al.~\cite{Theofanous1969}, where some level of thermodynamic non-equilibrium is introduced, showing better agreement with flow rate measurements for the flashing flow through a sharp-edged nozzle using the equilibrium model. Narayanan~\cite{Narayanan2021} adopted thermal-equilibrium models by Singhal~\cite{Singhal2002} and Yuan~\cite{Weixing2001}, also based on the RP equation, to analyze the flashing flow through an orifice and a converging-diverging nozzle. After calibrating the model coefficients and implementing a variable saturation pressure and latent heat modeling in the case of the nozzle flow, a good agreement with measurements was found for the flow rate through the orifice and for the streamwise profiles of pressure and vapor production along the nozzle. In the work of Xu et al.~\cite{Xu2022}, the standard model of Zwart et al.~\cite{Zwart2004} was compared to the modified version by Zhang et al.~\cite{Zhang2010}, accounting for thermodynamic non-equilibrium, in the simulation of cavitation and flash-boiling in a nozzle flow. Improved predictions of the discharge coefficient for the liquid phase was reported for the modified model at operating conditions leading to the onset of flashing due to the over-predicted vaporization given by the standard model.

Examples of simulations of flashing flows with full non-equilibrium models can be found in the work of Liao and Lucas~\cite{Liao2015} and Janet et al.~\cite{Janet2015}, where a two-fluid model was employed to model flash-boiling in a converging-nozzle. Using a multi-fluid model, where a separate set of governing equations are solved for the liquid and vapor phase, thermodynamic non-equilibrium between phases can be modeled explicitly along with other non-equilibrium effects such as the relative velocity between the liquid and the vapor flow. Despite the complexity of this technique, requiring the setup of several models for the so-called interphase transfer phenomena, the results show similar accuracy to equilibrium models for the streamwise profiles of pressure and vapor production along the nozzle, whereas the radial profiles of vapor fraction are found in poor agreement with measurements. An alternative to the complexity of two-fluids models to account for the thermodynamic non-equilibrium associated with flash-boiling is given by the Homogeneous Relaxation Model (HRM), where non-equilibrium is introduced empirically in the interphase mass transfer by a relaxation equation~\cite{Zapolski1996}. The HRM model was used by~\cite{Lyras2018} in the simulation of the flashing-flow through a sharp-edged orifice, showing good agreement with measurements for the computed mass flow rate for varying inlet pressures. However, profiles of the computed vapor fraction were not reported in the study.

In the present work, where a CFD model is employed for the first time in the simulation of the flashing-flow inside the pMDI, the model proposed by Kunz~\cite{Kunz2000} for cavitating flows is employed. The model is part of the standard cavitation models available in OpenFOAM along with those of Merkle~\cite{Merkle1998} and Schnerr-Sauer~\cite{Schnerr2001}, and it was selected because of the reduced number of input parameters and improved numerical stability compared to the two others. In the Kunz model, the source terms in the continuity and volume fraction equations~(\ref{eqn:continuityEquation}) and~(\ref{eqn:volumeFractionEquation}) assume the following form:

\begin{eqnarray}\label{eqn:continuityVolumeFractionPhaseChangeSourceTerms}
    S_m & = & \left(\dot{m}^+-\dot{m}^-\right)\left(\frac{1}{\rho_l}-\frac{1}{\rho_v}\right) \\
    S_{\alpha_l} & = & \left(\dot{m}^+-\dot{m}^-\right)\frac{1}{\rho_l} 
\end{eqnarray}

where $\dot{m}^-$ and $\dot{m}^+$ are the evaporation and condensation rates associated with the flash-boiling, defined as follows:

\begin{eqnarray}\label{eqn:evaporationCondensationRates}
    \dot{m}^- & = & \frac{C_v\rho_v\alpha_l\text{min}\left(0,p-p_{\text{sat}}\right)}{1/2\rho_lU^2_{\infty}t_{\infty}}\label{eqn:KunzEvaporationRate} \\
    \dot{m}^+ & = & \frac{C_c\rho_v\alpha_l^2\left(1-\alpha_l\right)}{t_{\infty}}\label{eqn:KunzCondensationRate}
\end{eqnarray}

where $C_v$ and $C_c$ are the empirical constants of the model and $U_{\infty}$ and $t_{\infty}$ the characteristic velocity and time scales of the main flow. A temperature dependent saturation pressure is implemented in the present model and results compared to those obtained with a constant saturation pressure calculated at room temperature to study the effect on the metered discharge. The source term $S_T$ of the energy equation~(\ref{eqn:energyEquation}) is finally defined as follows:

\begin{equation}\label{eqn:energyPhaseChangeSourceTerm}
    S_T=\left(\dot{m}^+-\dot{m}^-\right)\left(\frac{\alpha_l}{c_{v,l}}-\frac{\alpha_v}{c_{v,v}}\right)H_f
\end{equation}

where $c_{v,l}$ and $c_{v,v}$ are the specific isochoric heat capacity of the liquid and vapor phases, respectively, and $H_f$ the latent heat of vaporization of the liquid propellant, calculated at the mixture temperature. It is important to stress the complexity of flash-boiling is such that any given model carries a certain amount of empiricism, and tuning of the empirical constants cannot be generally avoided to achieve accurate predictions for specific applications. In this study, a sensitivity analysis to the evaporation and condensation coefficients of the Kunz model is presented in Section~\ref{subsection:results:flashBoilingCalibration}.

\subsection{Computational grid}\label{subsection:computationalGrid}
\begin{figure}
\centering
\includegraphics[width=1.0\textwidth,trim={0 0 0 0},clip]{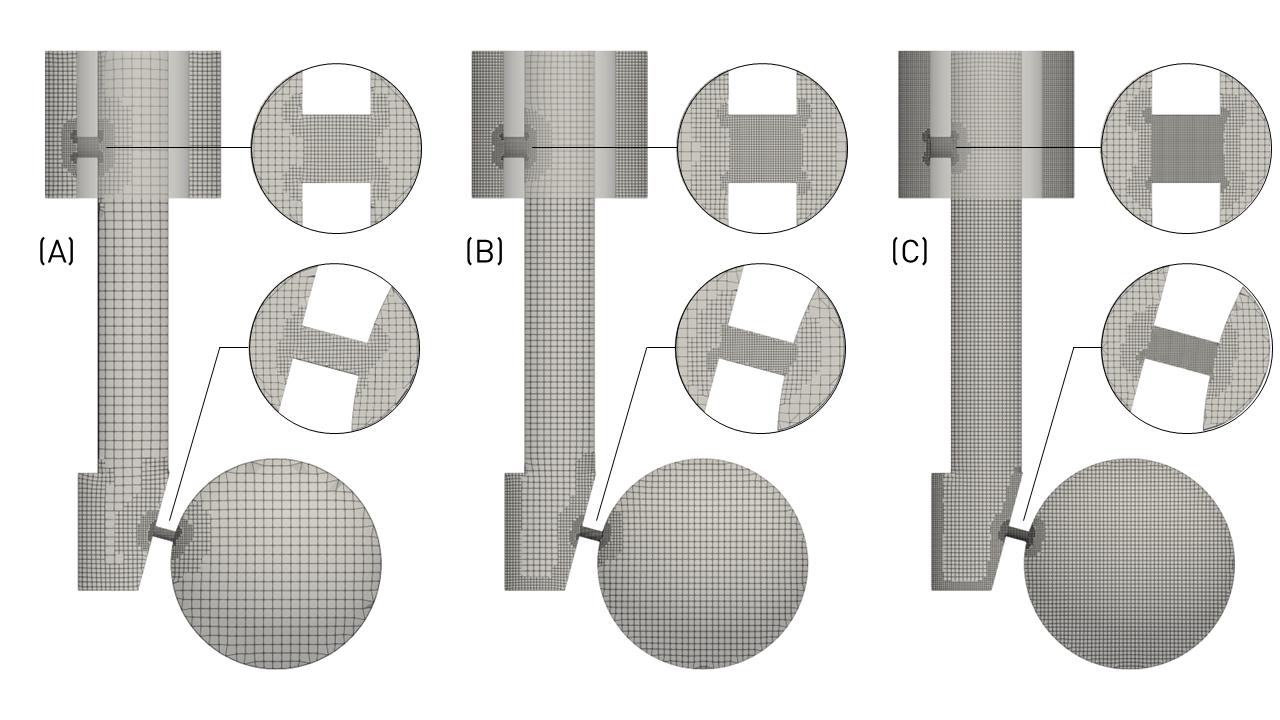}
\caption{Computational grids employed in the calculations: (a) coarse grid resolution, (b) medium grid resolution, (c) fine grid resolution. A close-up view of the grid resolution in the valve orifice and in the nozzle orifice is shown in the circles.}
\label{fig:computationalModel:computationalGridsOverview}
\end{figure}

The computational domain of the pMDI model extends from the metered valve to the nozzle orifice and includes the valve orifice and the expansion chamber, consisting of the valve stem and sump regions (see Fig.~\ref{fig:introduction:deviceGeometry:deviceGeometry}). A spherical plenum connected to the outlet section of the nozzle orifice is also included in the model to allow for the mixture flow to exit the device freely. Hexahedral-dominant grids built with the OpenFOAM$^\text{\textregistered}$ utility \texttt{snappyHexMesh} are employed for the discretization of the computational domain. In Fig.~\ref{fig:computationalModel:computationalGridsOverview}, a cross section of the computational grids with increasing resolution employed in the simulations is shown. In the coarse grid of panel~\ref{fig:computationalModel:computationalGridsOverview}(a), a minimum and maximum resolution of \SI{250}{\micro\meter} and about \SI{30}{\micro\meter}, respectively, are employed. The grid resolution is increased to a maximum and minimum of \SI{20}{\micro\meter} and \SI{167}{\micro\meter} and to \SI{13}{\micro\meter} and \SI{111}{\micro\meter} in the medium and fine grid, respectively. For the case of turbulent calculation, an additional grid with increased grid resolution at the wall is created, starting from the medium grid resolution. In this grid, three boundary layer cells are added at the wall with increased resolution in the wall-normal direction, ranging from about \SI{20}{\micro\meter} in the stem region to about \SI{5.5}{\micro\meter} in the nozzle orifice region. The total cell count of the computational grids is about 70k, 180k and 580k cells for the coarse, medium and fine grid, respectively.

\subsection{Numerical method}\label{subsection:numericalMethod}
The OpenFOAM$^\text{\textregistered}$ solver \texttt{compressibleInterFoam} for compressible, non-isothermal immiscible fluids, adopting the Volume-Of-Fluid interface capturing approach~\cite{Nichols1975} for the representation of the liquid/vapor mixture flow, is employed in the calculations. The model equations~(\ref{eqn:continuityEquation})-(\ref{eqn:volumeFractionEquation}) are discretized on the computational grid using second-order accurate schemes in space. For the convective and diffusive transport terms of the momentum equation and energy equations~(\ref{eqn:momentumEquation}) and~(\ref{eqn:energyEquation}), a linear-upwind and a central-linear scheme are used, respectively. A cell-limited Green-Gauss based reconstruction of cell-centered gradients is used in the linear-upwind scheme to improve numerical stability. For the convective term of the volume fraction equation~(\ref{eqn:volumeFractionEquation}), the scheme of Van Leer~\cite{VanLeer1977} is adopted. The numerical integration in time of the model equations is performed using a first-order implicit Euler scheme and an adjustable time-stepping technique, where a variable times-step size $\Delta t$ is calculated at run-time based on a maximum Courant number $\text{Co}=U\Delta t/\Delta x$ of 4 inside the liquid and vapor regions and of 2 at the interface between phases, where $U$ and $\Delta x$ are local mixture velocity and grid resolution, respectively. The average time-step size during a metered discharge calculation is about $1.5\times 10^{-4}$ \SI{}{\ms}, thus an accurate integration in time of the flow governing equations is expected despite the first-order scheme employed in the simulations. The computational effort for the simulation of a single metered discharge of 350 \SI{}{\ms} ranges from 5 hours for the coarse grid to 27 hours for the fine grid in parallel execution of the code on 192 and 384 cores, respectively.

\section{Results and discussion}\label{section:results}
The results of the present work are presented starting from a qualitative analysis of the simulated metered discharge, followed by an assessment of results sensitivity to the main modeling parameters, such as the flash-boiling model coefficients, the grid resolution, temperature dependent properties of the mixture and turbulence modeling. The measurements of mixture density and liquid/vapor flow rates at the nozzle orifice reported by Mason et al.~\cite{MasonSmith2017} for the same device geometry and operating conditions are mainly adopted to validate the numerical solution. Other experimental studies for devices with similar geometrical parameters and/or conditions of the metered discharge are employed to evaluate more general behavior of the present numerical model.

All the calculations are started from the same initial state, where the propellant is virtually stored in the metering chamber volume by setting the local liquid volume fraction to $\alpha_l=1$. The initial propellant pressure and temperature in the metering chamber are set to 5.9 bar and \SI{20}{\celsius}. In the remaining of the computational domain, i.e. the valve orifice, expansion chamber, nozzle orifice and outlet plenum, pressure and temperature are initialized to 1 bar and \SI{20}{\celsius} and the liquid volume fraction to $\alpha_l=0$, corresponding to pure vapor. Such condition does not represent the initial state of the actual device, which is initially filled with air. Nonetheless, the displacement of air by the incoming mixture flow is very rapid and usually neglected in system models as well~\cite{Clark1991}. The modeling of the device actuation, where the metering chamber moves initially with respect to the valve stem until the valve orifice is in full contact with the propellant, is also neglected in the present study, where no partialization of the valve orifice is considered.

The profiles of monitored variables during the simulated metered discharge are obtained by averaging the numerical solution in the volume of the device components with a sampling rate of $1\times 10^{-2}$ \SI{}{\ms}. In the case of the nozzle orifice, where high-scattering is observed due to the local high-speed flow, a moving average is applied to extracted profiles using a window of $10^3$ samples.

\subsection{Qualitative analysis of simulated metered discharge and comparison with experimental visualizations}
In this section, a qualitative analysis of the simulated metered discharge is carried out via the contour maps of liquid volume fraction. The fine grid resolution is employed in the calculation and the coefficients of the flash-boiling model ($C_v=0.0375$ and $C_c=240$) set from the calibration study against measured data presented in Section~\ref{subsection:results:flashBoilingCalibration}. Constant mixture properties, including the saturation pressure, are adopted in the calculation and the effect of latent heat on the energy equation neglected at this stage. As it will be shown later in the paper, the qualitative results presented here are not altered significantly by the main modeling parameters. Therefore, the analysis can be assumed of general validity in the context of the present model. 

\begin{figure}
\centering
\includegraphics[width=1.0\textwidth,trim={0 0 0 0},clip]{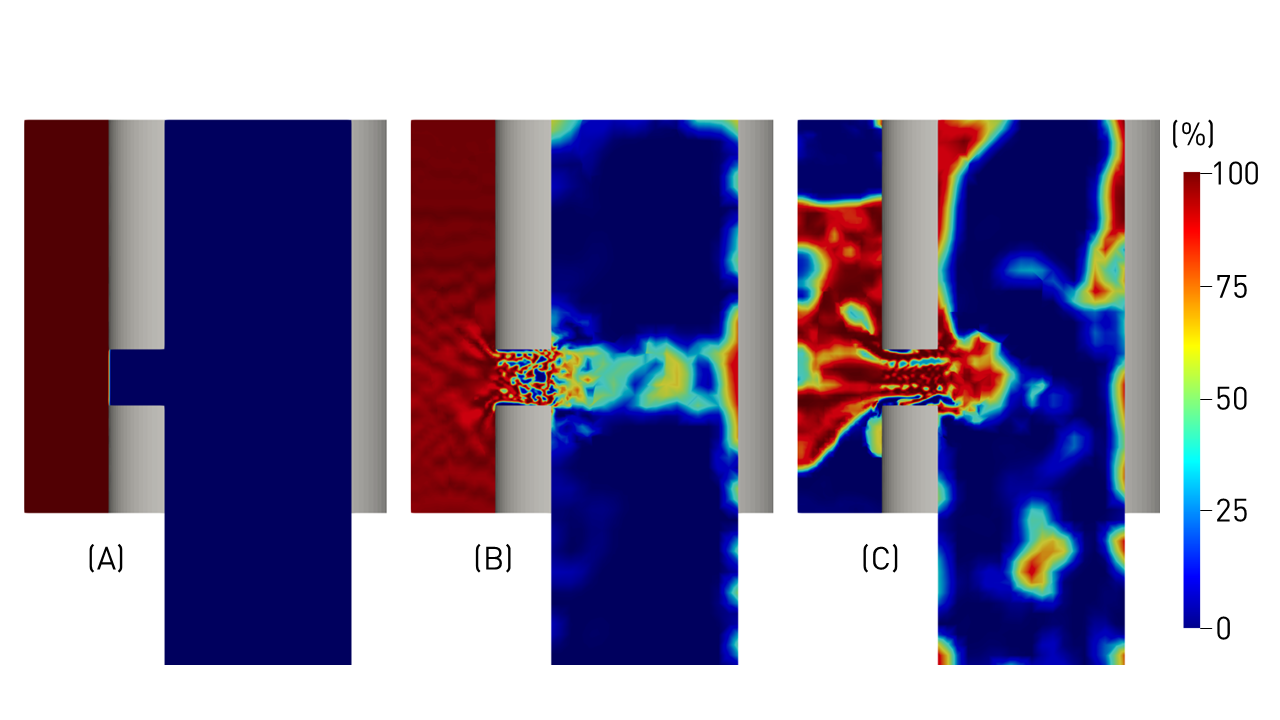}
\caption{Temporal evolution of the liquid fraction in the valve orifice region during the startup phase of the metered discharge: (a) $t$=0 \SI{}{\ms}, (b) $t$=0.3 \SI{}{\ms}, (c) $t$=30 \SI{}{\ms}.}
\label{fig:results:qualitativeAnalysis:liquidFractionTimeSequenceValveOrifice}
\end{figure}

In Fig.~\ref{fig:results:qualitativeAnalysis:liquidFractionTimeSequenceValveOrifice}, the evolution of the liquid fraction field in the valve orifice region is shown at the early stage of the metered discharge. The contour map at $t=0$ \SI{}{\ms} in panel (a) shows the initial condition of the calculation, where the propellant is stored inside the metered chamber above the saturation pressure and exposed instantaneously to the valve orifice volume, set to atmospheric pressure. After few instants ($t=0.3$ \SI{}{\ms}), the abrupt formation of vapor can be observed in the orifice volume in panel (b) due to the onset of the flash-boiling, generating a two-phase jet flow from the valve orifice towards the expansion chamber. The impingement of the jet flow on the opposite side of the valve stem can be also noticed, with partial condensation of the mixture at the stagnation point and the subsequent formation of a liquid film at the wall due to the effect of surface tension. The same panel also shows the perturbations associated with flash-boiling in the valve orifice traveling upstream into the metering chamber volume, triggering the local phase-change process and the initial reduction of the liquid volume fraction. At $t=30$ \SI{}{\ms} from the start of the metered discharge, the intensity of the jet flow is greatly reduced, as shown in Fig.~\ref{fig:results:qualitativeAnalysis:liquidFractionTimeSequenceValveOrifice}(c), and large scale structures of liquid and vapor are formed and released inside the expansion chamber. Because of the jet impingement at the wall, some structures travel directly down the valve stem towards the sump region whereas some others moves initially upstream, causing some liquid accumulation in the upper region of the stem. Large voids occupied by the vapor phase are found at this stage inside the metering chamber volume.

\begin{figure}
\centering
\includegraphics[width=1.0\textwidth,trim={0 0 0 0},clip]{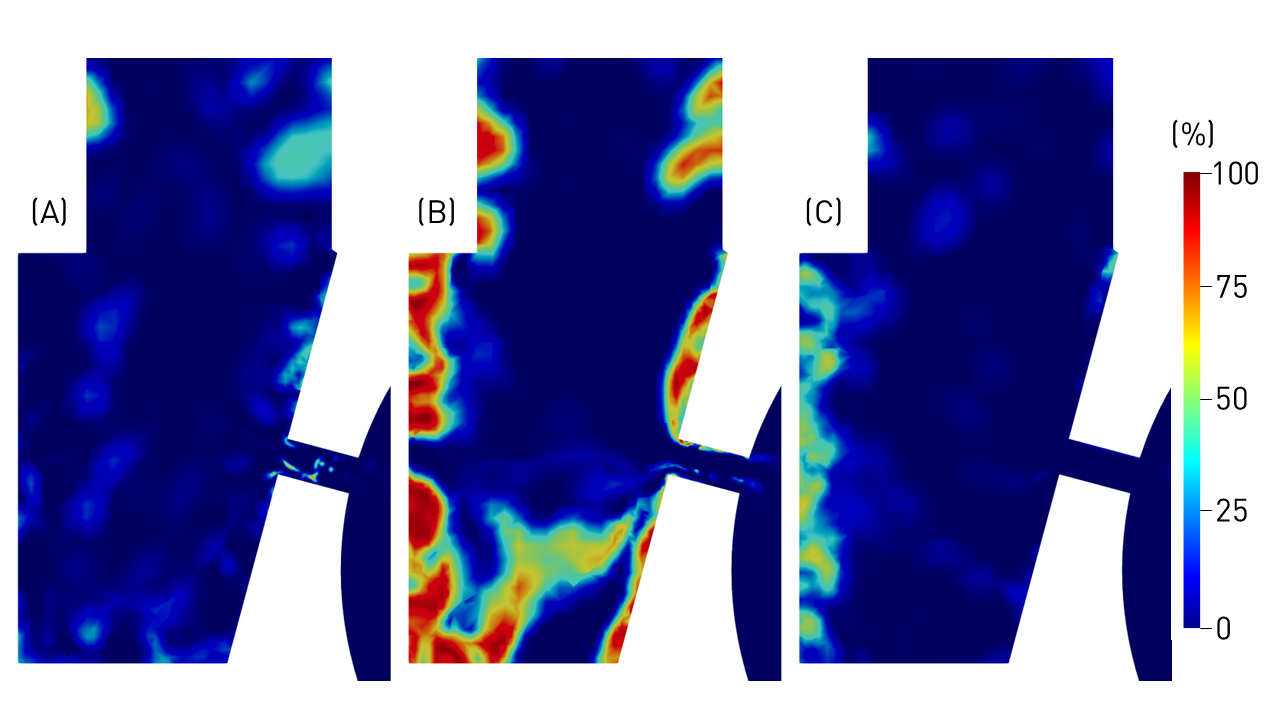}
\caption{Temporal evolution of the liquid fraction in the sump region from the startup to the decay phase of the metered discharge: (a) $t$=1.5 \SI{}{\ms}, (b) $t$=30 \SI{}{\ms}, (c) $t$=150 \SI{}{\ms}.}
\label{fig:results:qualitativeAnalysis:liquidFractionTimeSequenceSump}
\end{figure}

The evolution of the liquid fraction in the sump region from the initial to the final stage of the metered discharge is presented in Fig.~\ref{fig:results:qualitativeAnalysis:liquidFractionTimeSequenceSump}. In the startup phase ($t=1.5$ \SI{}{\ms}), shown in panel (a), high vapor-quality structures can be seen entering and filling up the sump volume, as well as structures with high liquid volume fraction already entering the nozzle orifice. At $t=30$ \SI{}{\ms}, liquid accumulation in the sump volume due to flow recirculation is observed in Fig.~\ref{fig:results:qualitativeAnalysis:liquidFractionTimeSequenceSump}(b), along with a thick liquid film forming around the inlet section of the nozzle orifice. In the decay phase of the metered discharge ($t=150$ \SI{}{\ms}), shown in Fig.~\ref{fig:results:qualitativeAnalysis:liquidFractionTimeSequenceSump}(c), a region of stagnant liquid in the back wall of the sump is found. At this stage, only high vapor quality structures can be seen entering the nozzle orifice.

\begin{figure}
\centering
\includegraphics[width=1.0\textwidth,trim={0 0 0 0},clip]{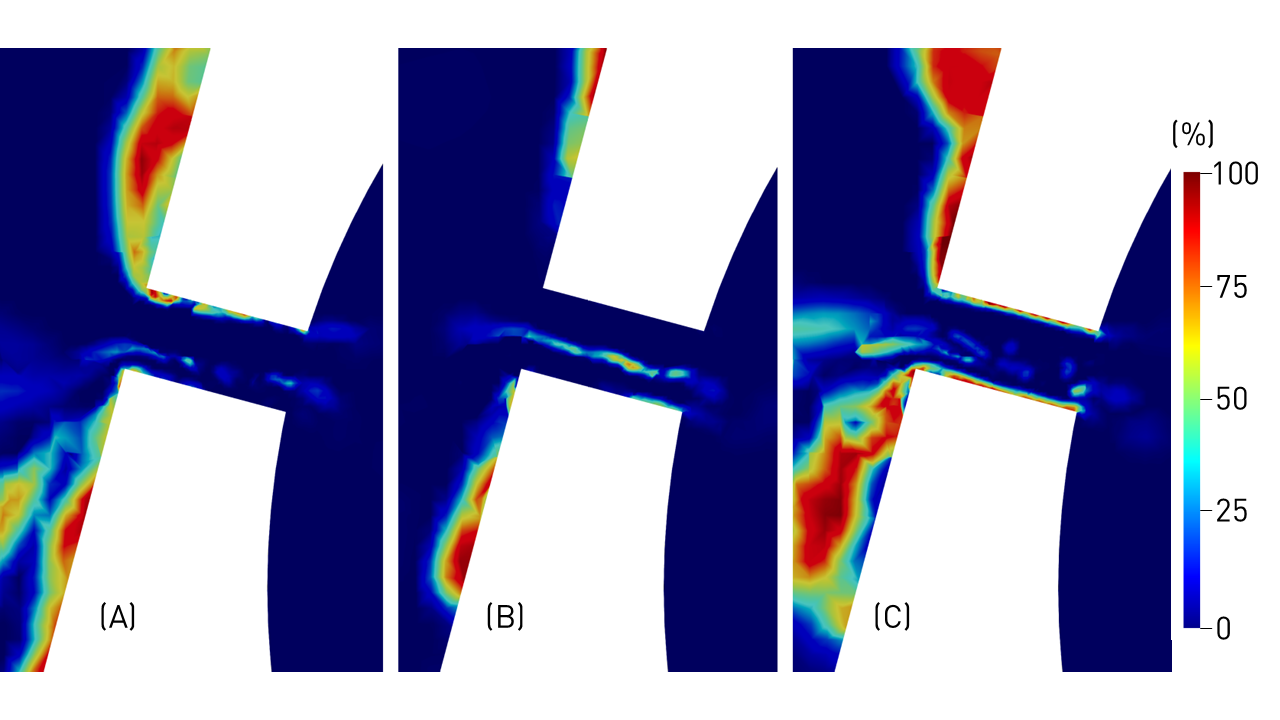}
\caption{Temporal evolution of the liquid fraction in the nozzle orifice region in the intermediate phase of the metered discharge: (a) $t$=30 \SI{}{\ms}, (b) $t$=60 \SI{}{\ms}, (c) $t$=90 \SI{}{\ms}.}
\label{fig:results:qualitativeAnalysis:liquidFractionTimeSequenceNozzleOrifice}
\end{figure}

A close-up view of the mixture flow in the nozzle orifice is presented in Fig.~\ref{fig:results:qualitativeAnalysis:liquidFractionTimeSequenceNozzleOrifice}. In this region, the liquid and vapor structures experience a strong acceleration by moving from the low-speed, recirculating flow in the sump volume to the high-speed flow inside the nozzle. Here, the liquid structures also experience significant stretching, as shown by the elongated shape appearing at $t=60$ \SI{}{\ms} in Fig.~\ref{fig:results:qualitativeAnalysis:liquidFractionTimeSequenceNozzleOrifice}(b). Fig.~\ref{fig:results:qualitativeAnalysis:liquidFractionTimeSequenceNozzleOrifice}(a) and Fig.~\ref{fig:results:qualitativeAnalysis:liquidFractionTimeSequenceNozzleOrifice}(c) also show the formation of a liquid film on the wall surface of the nozzle orifice due to the ingestion of the liquid film formed on the front wall of the sump, co-existing with the liquid and vapor structures traveling into the core region of the orifice.

\begin{figure}
\centering
\begin{subfigure}{1.0\textwidth}
\centering
\includegraphics[width=1.0\textwidth,trim={30 110 30 110},clip]{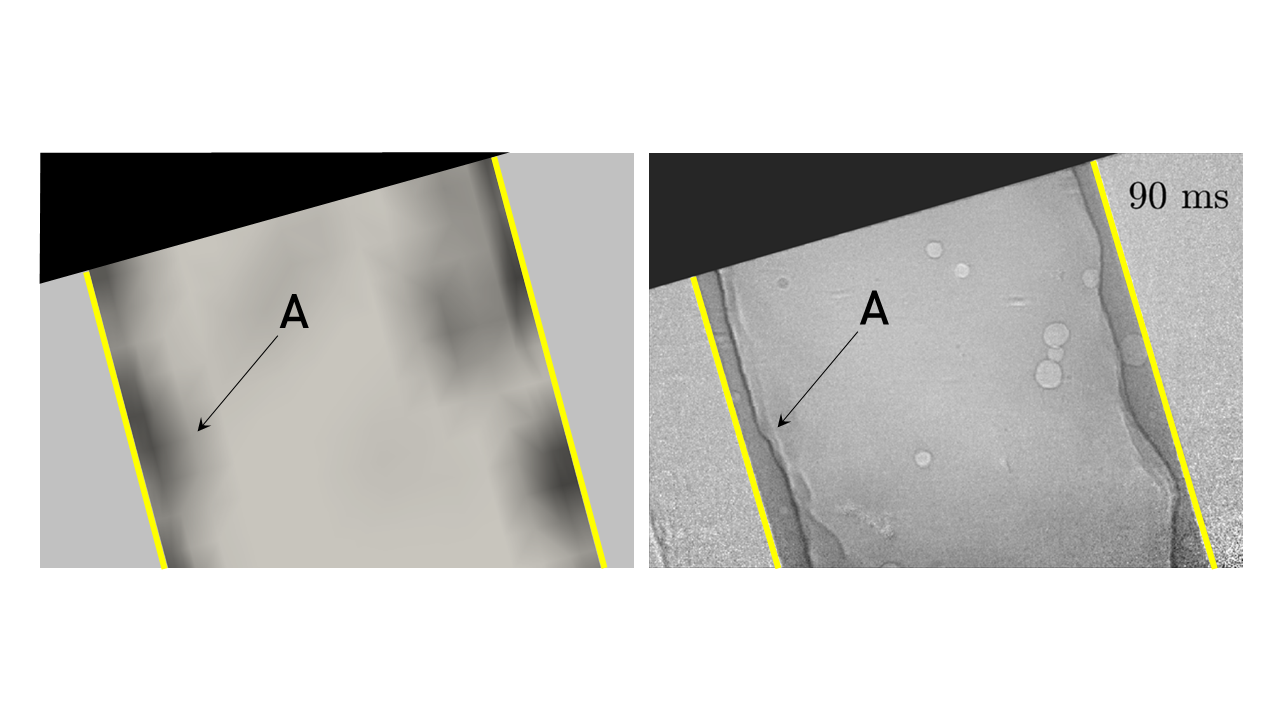}
\end{subfigure}
\begin{subfigure}{1.0\textwidth}
\centering
\includegraphics[width=1.0\textwidth,trim={30 110 30 110},clip]{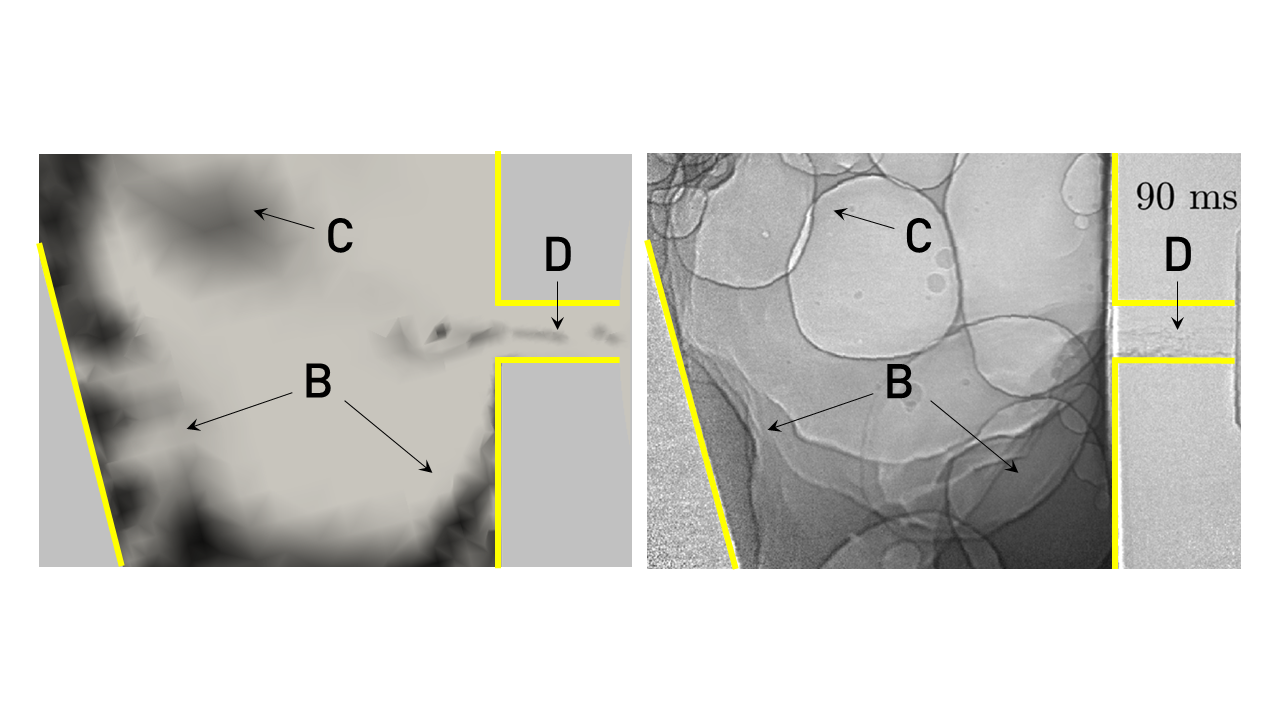}
\end{subfigure}
\caption{Comparison of computed liquid volume fraction field (left) with X-ray visualizations (right) in the valve stem (top) and sump (bottom) regions (X-ray images are extracted from the original movie courtesy of Mason-Smith et al.~\cite{MasonSmith2017}): (A) wall-film formation in the stem, (B) formation of large vapor structures in the core of the sump, (C) liquid recirculation and accumulation in the bottom part of the sump, (D) liquid ligaments undergoing acceleration and stretching in the nozzle orifice.}
\label{fig:results:qualitativeAnalysis:XRayComparison}
\end{figure}

In Fig.~\ref{fig:results:qualitativeAnalysis:XRayComparison}, a qualitative comparison of the present numerical solution with the experimental X-Ray visualizations of Mason-Smith et al.~\cite{MasonSmith2017} is presented to assess the capability of the CFD model to provide a realistic representation of the metered discharge. In the picture, the liquid volume fraction is shown in grayscale for the sake of comparison with the X-Ray images. Although the grid resolution is not sufficient to capture its actual thickness, the wall liquid-film predicted by the CFD model on the valve stem surface (detail A in the figure) is also present in the experimental visualizations. On the other hand, the similarity of the numerical solution with the X-Ray visualization in the sump region is quite remarkable. The liquid accumulation in the bottom part of the sump shown by the CFD model (detail B in the figure), particularly in the lower-right corner, can be clearly seen in the experiment. Moreover, even though the VOF model and the grid resolution employed in the computation are unable to properly resolve the vapor-liquid interface, the presence of large-scale vapor structures is also seen in the X-Ray image (detail C in the figure). The liquid ligaments formed in the nozzle (detail D in the figure) are also highlighted clearly by the X-Ray visualizations.

\begin{figure}
\centering
\begin{subfigure}{0.495\textwidth}
\centering
\includegraphics[width=\textwidth,trim={0 0 0 0},clip]{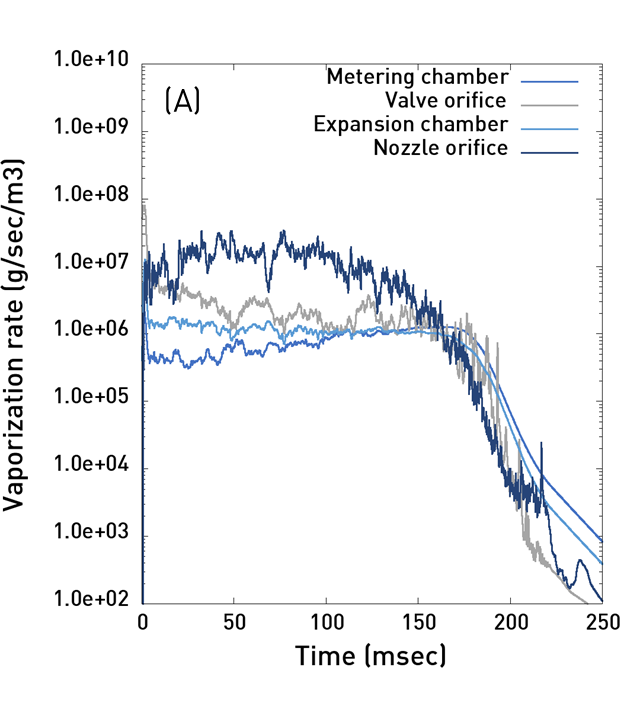}
\end{subfigure}
\hfill
\begin{subfigure}{0.495\textwidth}
\centering
\includegraphics[width=\textwidth,trim={0 0 0 0},clip]{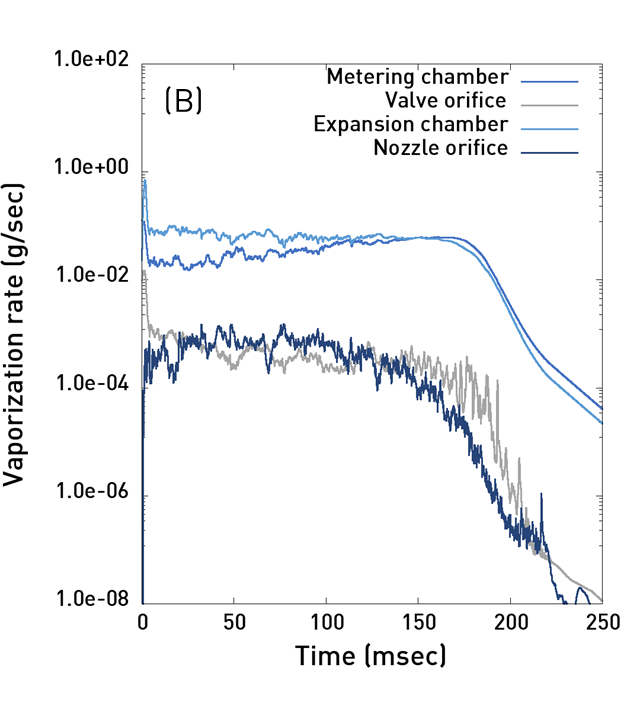}
\end{subfigure}
\caption{Vaporization rate associated with the main device components: (a) volume-averaged vaporization rate, (b) volume-integrated vaporization rate.}
\label{fig:results:qualitativeAnalysis:vaporisationRateComparison}
\end{figure}

The intensity of the flash-boiling and vapor production associated with device components is analyzed in Fig.~\ref{fig:results:qualitativeAnalysis:vaporisationRateComparison} by plotting the volume-averaged and volume-integrated source term of the liquid volume fraction equation~(\ref{eqn:volumeFractionEquation}) during the metered discharge. The volume-averaged vaporization rate, presented in panel (a), shows that in the very early stage of device operation ($t<3$ \SI{}{\ms}), the maximum intensity of flash-boiling takes place in the valve-orifice. However, as the mixture reaches the nozzle orifice, the intensity of the local flash-boiling is found an order of magnitude higher compared to the other device components due to the significant pressure reduction associated with the high-speed flow through the nozzle. On the other hand, the intensity of the flash-boiling in the valve orifice rapidly decreases reaching the same order of magnitude as in the expansion and metering chambers at $t\approx 50$ \SI{}{\ms}. However, the volume integral of the vaporization rate in panel (b) shows that the amount of vapor produced by the two chambers is significantly larger compared to the two orifices.

\subsection{Calibration of the flash-boiling model}\label{subsection:results:flashBoilingCalibration}
\begin{figure}
\centering
\begin{subfigure}{0.495\textwidth}
\centering
\includegraphics[width=\textwidth,trim={0 0 0 0},clip]{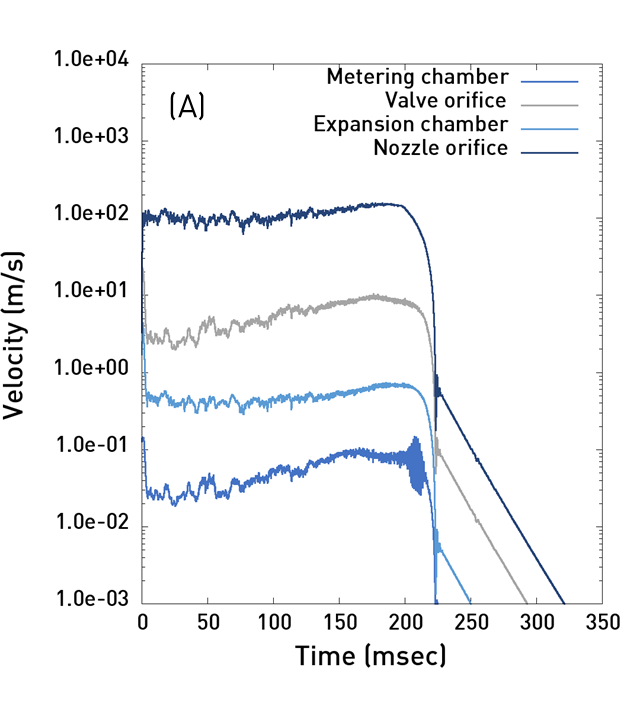}
\end{subfigure}
\hfill
\begin{subfigure}{0.495\textwidth}
\centering
\includegraphics[width=\textwidth,trim={0 0 0 0},clip]{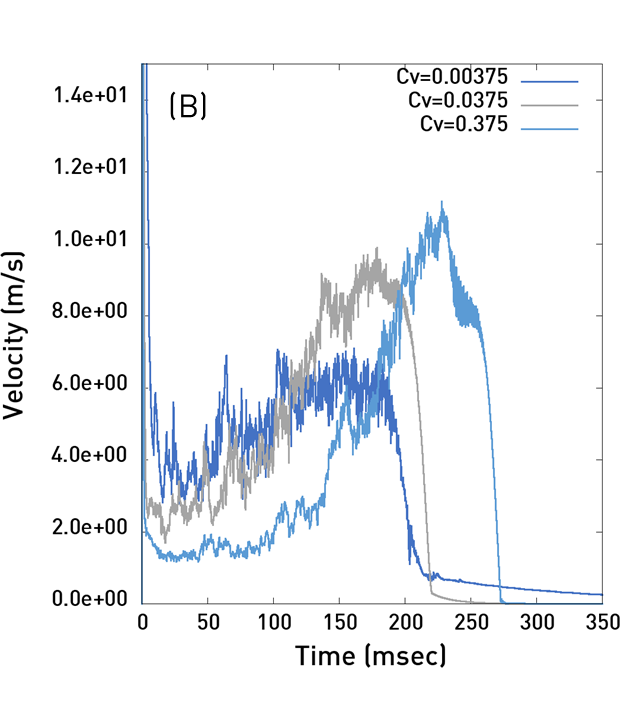}
\end{subfigure}
\caption{Mixture flow velocity profiles during metered discharge: (a) comparison among the main device components with $C_v=0.0375$, (b) effect of evaporation coefficient on velocity profiles in the valve orifice volume.}
\label{fig:results:flashBoilingCalibration:flowVelocityComparison}
\end{figure}

The Kunz model adopted in the present work to account for the flash-boiling phenomenon belongs to a class of phase-change models mainly developed for the simulation of external cavitating flows. In this type of flows, the characteristic velocity and temporal scales $U_{\infty}$ and $t_{\infty}$ in the model equations~(\ref{eqn:KunzEvaporationRate}) and~(\ref{eqn:KunzCondensationRate}) can be clearly associated with the main flow and thus easily identified. In the case of a two-orifice system as the pMDI, even if negligible mean flow speed is observed in the metering and expansion chamber (see Fig.~\ref{fig:results:flashBoilingCalibration:flowVelocityComparison}(a)), the characteristic flow scales can be either associated with the valve or the nozzle orifice. Despite the higher intensity of the flash-boiling taking place in the nozzle-orifice highlighted by Fig.~\ref{fig:results:qualitativeAnalysis:vaporisationRateComparison}(a), the valve-orifice flow has been selected to compute the characteristic scales, the focus of this work being directed towards the flow inside the pMDI and not on the generation of the spray and aerosol downstream of the nozzle orifice.  

The main calibration of the flash-boiling model coefficients presented here has been conducted using the coarse grid resolution and constant mixture properties. Further evaluations and adjustments have been performed during the sensitivity analysis to the main model parameters starting from the values identified at this stage of the model development, particularly for the case of variable saturation pressure (see Section~\ref{subsection:results:temperatureDependentProperties}). From a preliminary assessment of the mixture velocity in the valve orifice volume, the characteristic velocity and temporal scales have been set to $U_{\infty}=2.5$ m/s and $t_{\infty}=10^{-1}$ \SI{}{\ms} with $t_{\infty}=L_{vo}/U_{\infty}$, where $L_{vo}$ is the length of the valve orifice. Subsequently, the model coefficients have been varied starting from the standard values of $C_v=100$ and $C_c=100$ suggested in the work of Kunz~\cite{Kunz2000}, and the numerical solution evaluated by comparing the computed mixture density and liquid and vapor flow rates in the nozzle orifice with the measurements of Mason-Smith et al.~\cite{MasonSmith2017}. From preliminary calibration tests, it was found that the condensation coefficient has minimal effect on the numerical solution due to negligible condensation rate during the metered discharge where, in contrast to cavitating flows, no significant pressure recovery takes place. For this reason, once an optimal value was found for $C_c$ only the evaporation coefficient was further evaluated.

\begin{figure}
\centering
\begin{subfigure}{0.495\textwidth}
\centering
\includegraphics[width=\textwidth,trim={0 0 0 0},clip]{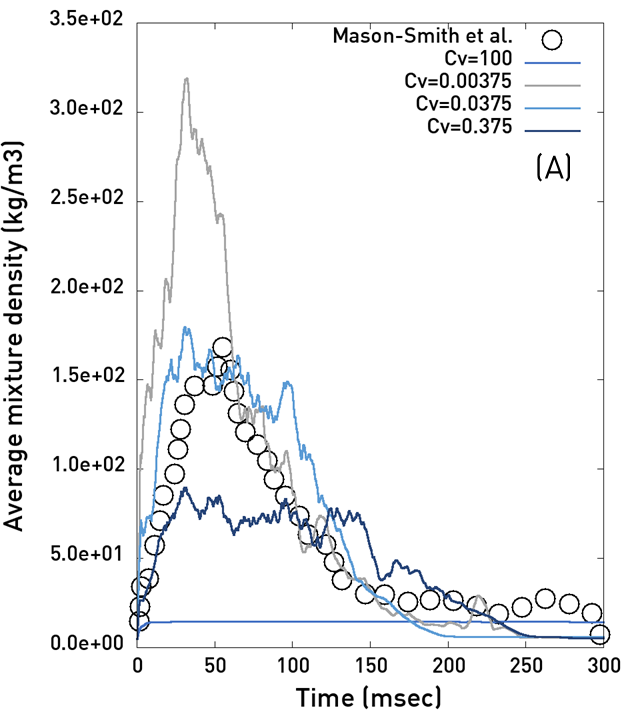}
\end{subfigure}
\hfill
\begin{subfigure}{0.495\textwidth}
\centering
\includegraphics[width=\textwidth,trim={0 0 0 0},clip]{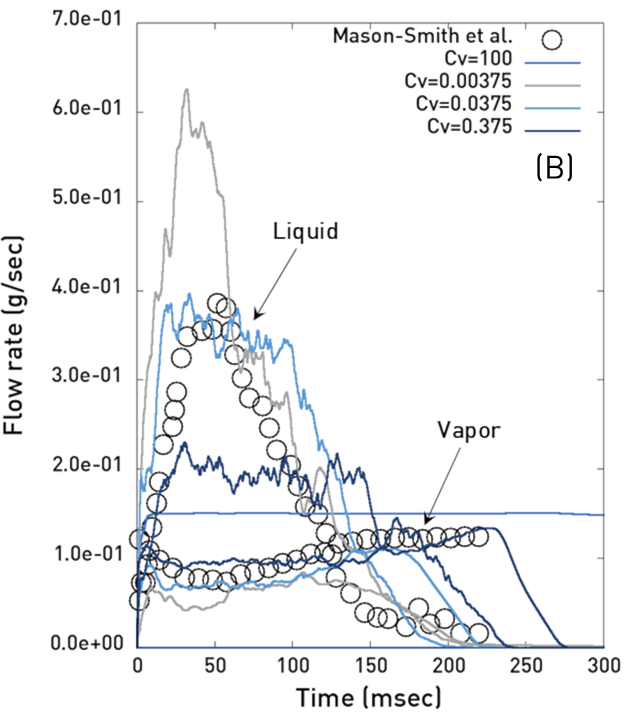}
\end{subfigure}
\caption{Comparison of computed mixture discharge profiles at the nozzle orifice with the measurements of Mason-Smith et al.~\cite{MasonSmith2017}: (a) mixture density, (b) liquid and vapor flow rates.}
\label{fig:results:flashBoilingCalibration:nozzleOrificeProfiles}
\end{figure}

Fig.~\ref{fig:results:flashBoilingCalibration:nozzleOrificeProfiles} shows the effect of varying the evaporation coefficient on the predicted mixture flow in the nozzle orifice. Using the standard coefficients, the flash-boiling is so intense to make the liquid evaporating entirely across the valve orifice, leading to a single-phase vapor flow in the expansion chamber and through the nozzle orifice. A very different value of $C_v=0.0375$, along with $C_c=240$, gives a good prediction of measured peak values of mixture density and liquid flow rate, whereas the experimental profiles are found overpredicted to some extent in the start-up and decay phases of the metered discharge. On the other hand, the profile of vapor flow rate is found mostly underestimated during the shot. Nonetheless, the overall agreement with the experimental profiles is remarkable given the level of complexity associated with the flash-boiling flow inside the pMDI.

By increasing/decreasing the evaporation coefficient by an order of magnitude, the peak values of mixture density decrease/increase by -50$\%$/+100$\%$ and those of liquid flow rate by -50$\%$/+50$\%$, respectively, whereas the maximum relative change in vapor flow rate is +36$\%$/-36$\%$. From the mixture density and liquid flow rate profiles, a more defined peak in the shot is observed with decreasing values of the evaporation coefficient, also leading to a better agreement with measurements in the decay phase of the metered discharge, while a non-monotonic behavior is found in the start-up phase. A significant effect on the duration of the shot, of about 50 \SI{}{\ms}, is also seen from the vapor flow rates profiles when moving from the optimal coefficient of $C_v=0.0375$ to $C_v=0.375$. It is also interesting to note that the predicted time of the peak in the shot is fairly insensitive to the calibration of the flash-boiling model and is in good agreement with the experiments. The {\em a-posteriori} evaluation of mixture velocity in the valve orifice volume, shown in Fig.~\ref{fig:results:flashBoilingCalibration:flowVelocityComparison}(b), reveals similar values up to the decay phase of the shot in the range $C_v=0.00375-0.0375$, whereas a more significant velocity variation is found for $C_v=0.375$. The use of a constant set of characteristic scales for the flash-boiling model thus seems a reasonable assumption.

\subsection{Grid sensitivity analysis}\label{subsection:results:gridSensitivity}
\begin{figure}
\centering
\begin{subfigure}{0.495\textwidth}
\centering
\includegraphics[width=\textwidth,trim={0 0 0 0},clip]{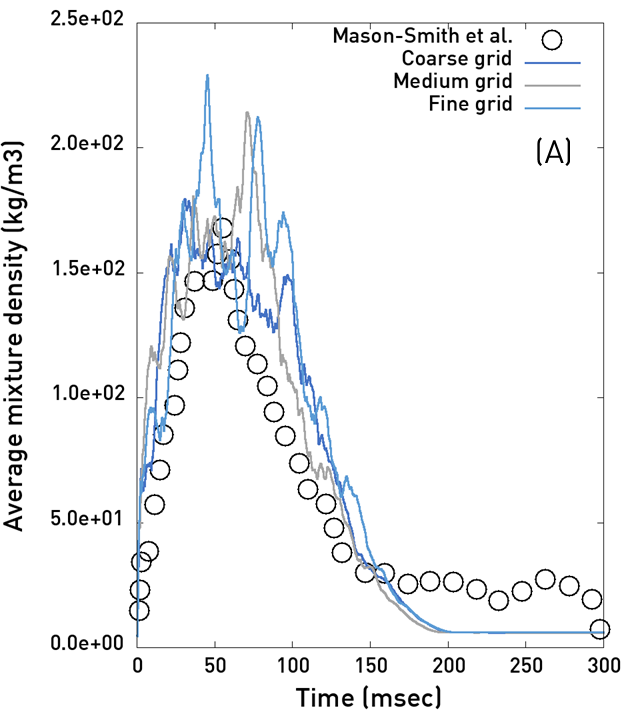}
\end{subfigure}
\hfill
\begin{subfigure}{0.495\textwidth}
\centering
\includegraphics[width=\textwidth,trim={0 0 0 0},clip]{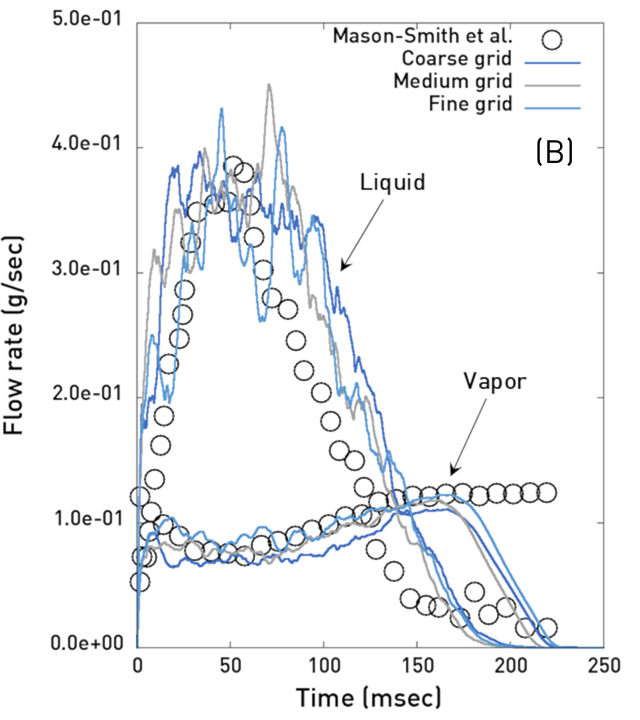}
\end{subfigure}
\caption{Impact of grid resolution on the computed mixture discharge profiles at the nozzle orifice: (a) mixture density, (b) liquid and vapor flow rates.}
\label{fig:results:gridSensitivity:nozzleOrificeProfiles}
\end{figure}

The effect of grid resolution on the computed metered discharge has been evaluated for the case of constant mixture properties and using the flash-boiling model coefficients identified in Section~\ref{subsection:results:flashBoilingCalibration}, giving the best agreement with experimental data ($C_v=0.0375$, $C_c=240$). In Fig.~\ref{fig:results:gridSensitivity:nozzleOrificeProfiles}, the computed profiles of mixture density and liquid and vapor flow rates at the nozzle orifice using the coarse, medium and fine grid are compared to the experimental profiles of Mason-Smith et al.~\cite{MasonSmith2017}. A marginal effect is observed on mixture density, where slightly higher fluctuations are found in peak values moving from the coarse to the medium and fine grid resolutions, without a significant change in the predicted profile. The grid resolution seems to have no effect on predicting the plateau of the mixture density observed in the experiments between 150 and 300 \SI{}{\ms}, followed by a sudden drop to the vapor density value. However, a notable improvement can be seen in the predicted profiles of liquid and vapor flow rates along the entire metered discharge when moving to the fine grid resolution, with a significant reduction of the overestimated liquid flow rate and, conversely, of the underpredicted vapor flow rate.

\begin{figure}
\centering
\begin{subfigure}{0.495\textwidth}
\centering
\includegraphics[width=\textwidth,trim={0 0 0 0},clip]{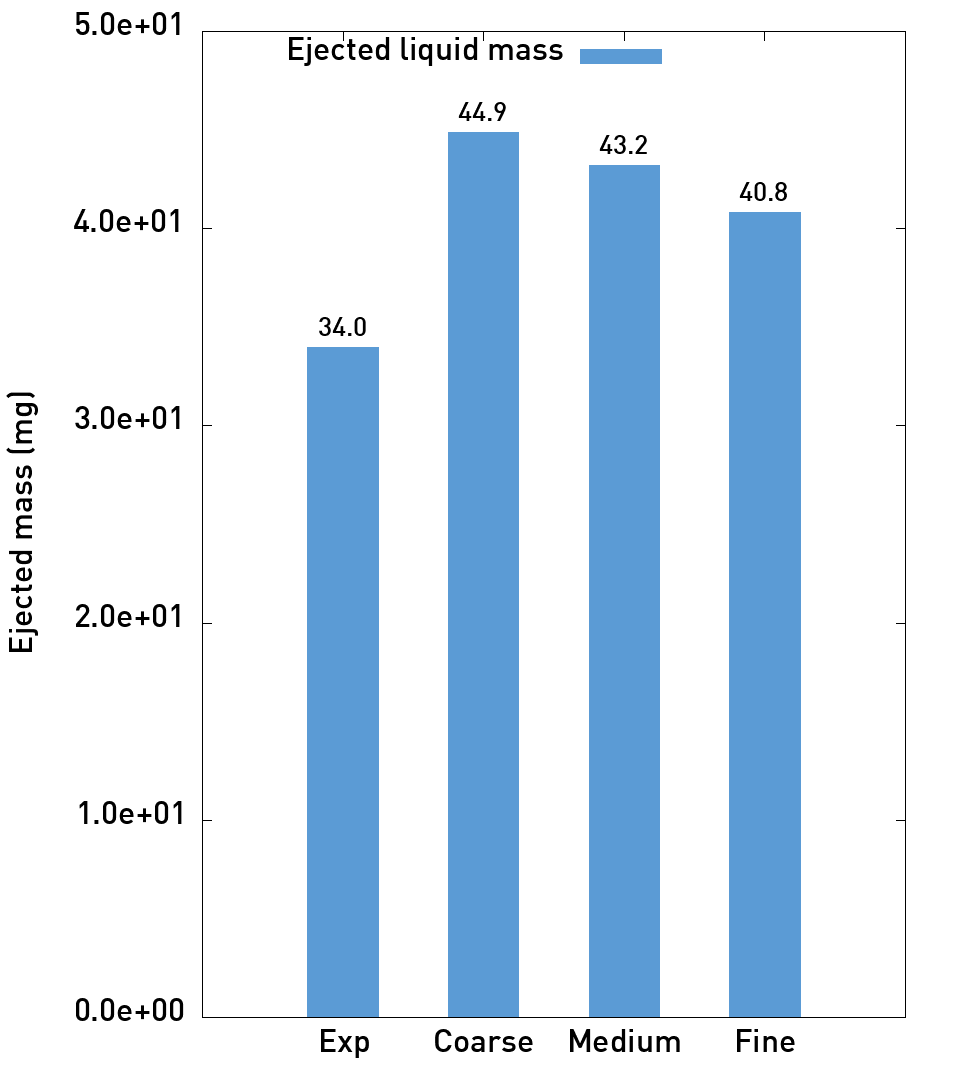}
\end{subfigure}
\hfill
\begin{subfigure}{0.495\textwidth}
\centering
\includegraphics[width=\textwidth,trim={0 0 0 0},clip]{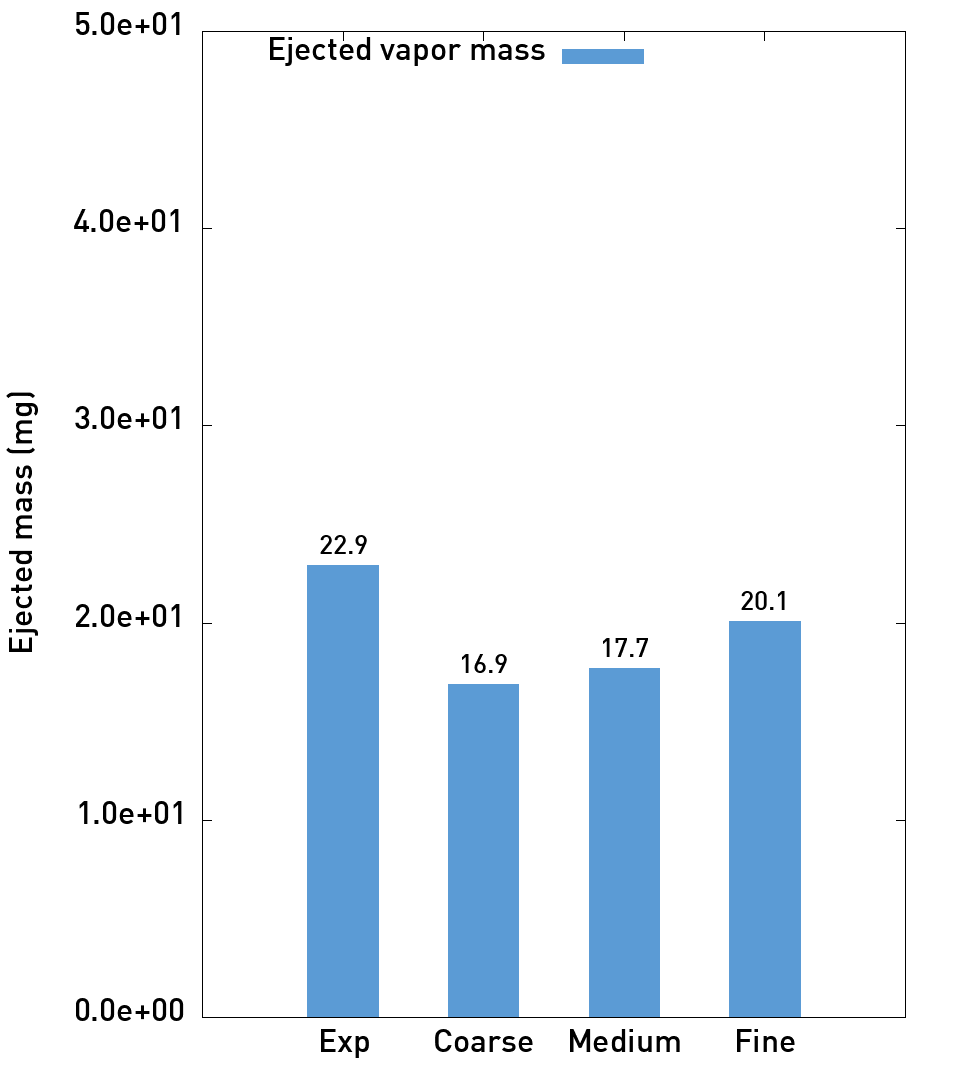}
\end{subfigure}
\caption{Impact of grid resolution on the computed ejected mass of liquid and vapor from the nozzle orifice against the experiments of Mason-Smith et al.~\cite{MasonSmith2017}: (a) liquid mass, (b) vapor mass.}
\label{fig:results:gridSensitivity:ejectedMassComparison}
\end{figure}

{The increased accuracy in the predicted flow rates at the nozzle orifice with increasing grid resolution is quantified and shown in Fig.~\ref{fig:results:gridSensitivity:ejectedMassComparison}, where the computed mass of liquid and vapor ejected during the metered discharge, given by the integral of the flow rate profiles, is compared to values measured in the experiments of Mason-Smith et al.~\cite{MasonSmith2017}. A systematic improvement is observed when moving from the coarse to the fine grid resolution, with the error in the computed ejected mass dropping from +32\% to +20\% for the liquid phase and from -26\% to -12\% for the vapor phase, respectively. The results also show the finest grid resolution employed in this study does not allow to reach a fully-converged prediction of the ejected mass, for which a variation of -5\% and +14\% is still observed between the medium and fine grids. Nonetheless, since the scope of the present work is the development of a numerical model for industrial applications, where the computational cost of the simulations must be limited, and no significant changes and shifts in the overall shape of the predicted profiles in Fig.~\ref{fig:results:gridSensitivity:nozzleOrificeProfiles} can be noticed, we deem the analysis presented in this section sufficient to assess the robustness of the numerical model.}

\begin{figure}
\centering
\begin{subfigure}{0.495\textwidth}
\centering
\includegraphics[width=\textwidth,trim={0 0 0 0},clip]{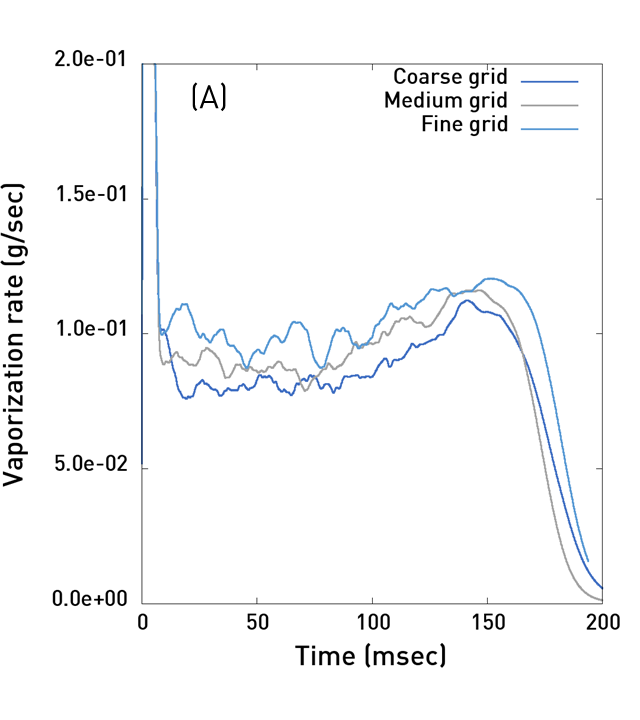}
\end{subfigure}
\hfill
\begin{subfigure}{0.495\textwidth}
\centering
\includegraphics[width=\textwidth,trim={0 0 0 0},clip]{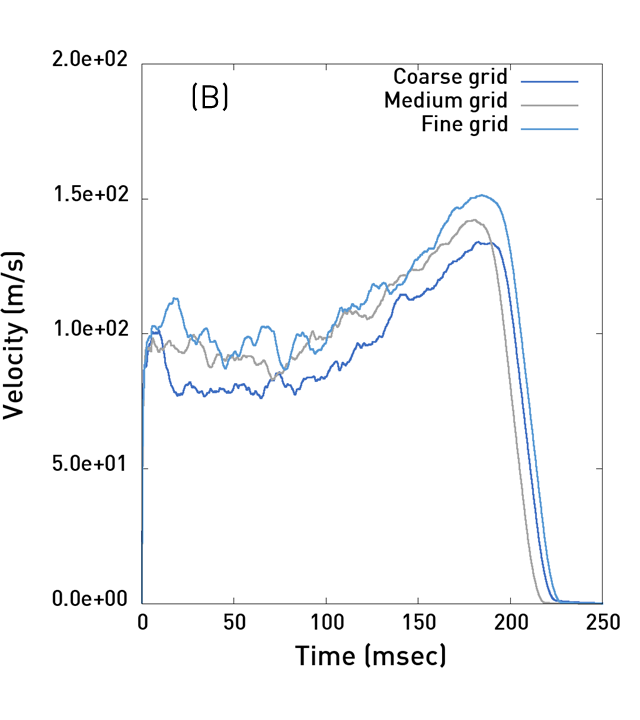}
\end{subfigure}
\caption{Effect of grid resolution on (a) global vaporization rate in the pMDI and on (b) mixture velocity in the nozzle orifice.}
\label{fig:results:gridSensitivity:nozzleOrificeVaporizationVelocityProfiles}
\end{figure}

The effect of grid resolution on the simulated metered discharge is further investigated in Fig.~\ref{fig:results:gridSensitivity:nozzleOrificeVaporizationVelocityProfiles} by the analysis of the global vaporization rate inside the pMDI and of the mixture velocity in the nozzle orifice. The increased mixture velocity with increasing grid resolution in Fig.~\ref{fig:results:gridSensitivity:nozzleOrificeVaporizationVelocityProfiles}(b) supports the improvement in the predicted vapor flow rate, but appears in contrast with the predicted liquid flow rate, which is found decreasing with increasing grid resolution. However, the profiles of global vaporization in Fig.~\ref{fig:results:gridSensitivity:nozzleOrificeVaporizationVelocityProfiles}(a) shows how vapor production, and thus liquid evaporation, increases with grid refinement. The analysis of mixture velocity in the valve orifice, not reported here for the sake of brevity, also reveals a partial increase with increasing grid resolution, leading to higher depressurization rate and thus to a more intense flash-boiling.

\subsection{Effect of temperature-dependent mixture properties}\label{subsection:results:temperatureDependentProperties}
During the metered discharge, a monotonic drop of pressure and temperature is experienced by the mixture in the metering chamber and, after the initial rise from room conditions, also in the expansion chamber~\cite{Clark1991}. The temperature drop is driven by the phase-change process as well as by the expansion of the vapor phase which results from the pressure drop across the valve and nozzle orifice and from the overall pressure reduction during the shot. Temperature measurements inside the pMDI are scarce and not well documented (see also the discussion in Section~\ref{subsection:results:mixtureTemperature}) but even considering the largest recorded temperature drops, the variation of mechanical and thermodynamic properties estimated from tabulated values for the HFA-134a, such as viscosity and heat capacity, are not significant. However, the saturation pressure of the propellant varies significantly with temperature, potentially influencing the flash-boiling process inside the device. In this section, the emphasis is thus set towards the analysis of the effect of variable saturation pressure on the simulated metered discharge.

\begin{figure}
\centering
\begin{subfigure}{0.495\textwidth}
\centering
\includegraphics[width=\textwidth,trim={0 0 0 0},clip]{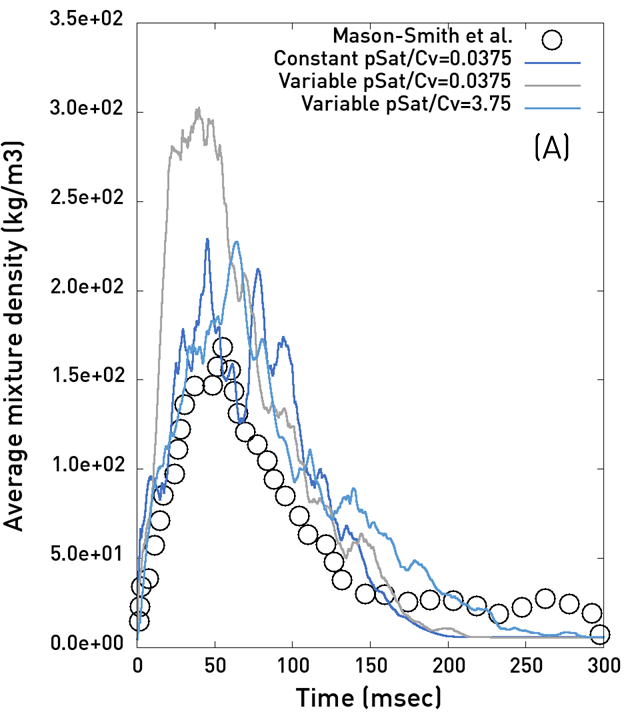}
\end{subfigure}
\hfill
\begin{subfigure}{0.495\textwidth}
\centering
\includegraphics[width=\textwidth,trim={0 0 0 0},clip]{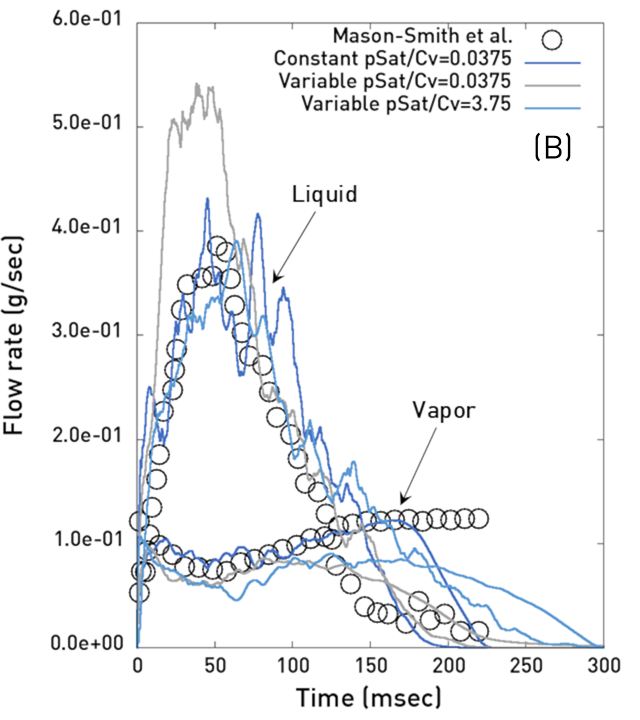}
\end{subfigure}
\caption{Effect of variable saturation pressure on computed mixture discharge profiles at the nozzle orifice: (a) mixture density, (b) liquid and vapor flow rates.}
\label{fig:results:temperatureDependentProperties:nozzleOrificeProfiles}
\end{figure}

The predicted mixture density and liquid/vapor flow rates using a constant and variable saturation pressure are compared in Fig.~\ref{fig:results:temperatureDependentProperties:nozzleOrificeProfiles}. The calculations are performed using the fine grid resolution and the standard set of evaporation and condensation coefficients ($C_v=0.0375$ and $C_c=240$). The latent heat contribution to the energy equation is still neglected at this stage and thus temperature changes are induced by pressure variations only. The mixture density profile shows a significant increase in the peak value in case of variable saturation pressure, which is found greatly overestimated compared to both the constant saturation pressure profile and measurements, suggesting a reduction in the intial vaporization rate associated with the flash boiling. This is also reflected in the liquid and vapor flow rates, which are found overestimated and underestimated, respectively, in the startup phase of the metered discharge ($t<60$ \SI{}{\ms}). In the decay phase of the shot, a significant drop in the vapor flow rate can be also noticed compared to the constant saturation profile and measured data.

\begin{figure}
\centering
\begin{subfigure}{0.495\textwidth}
\centering
\includegraphics[width=\textwidth,trim={0 0 0 0},clip]{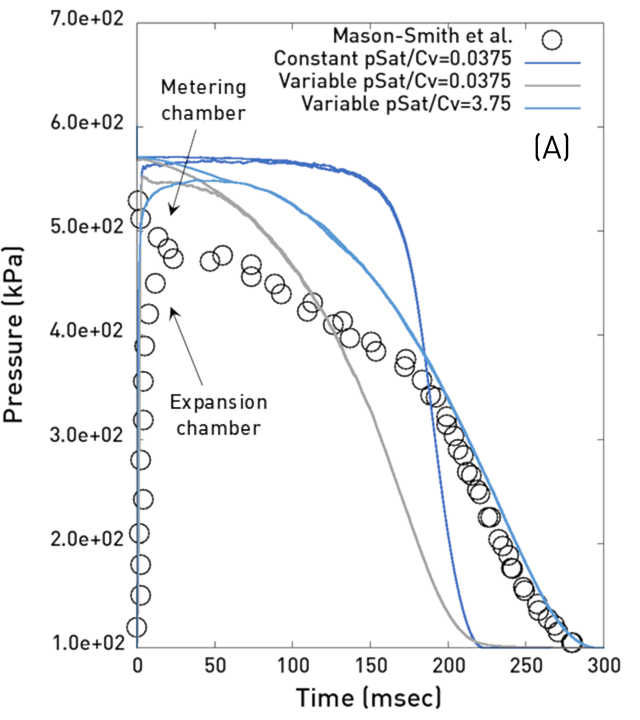}
\end{subfigure}
\hfill
\begin{subfigure}{0.495\textwidth}
\centering
\includegraphics[width=\textwidth,trim={0 0 0 0},clip]{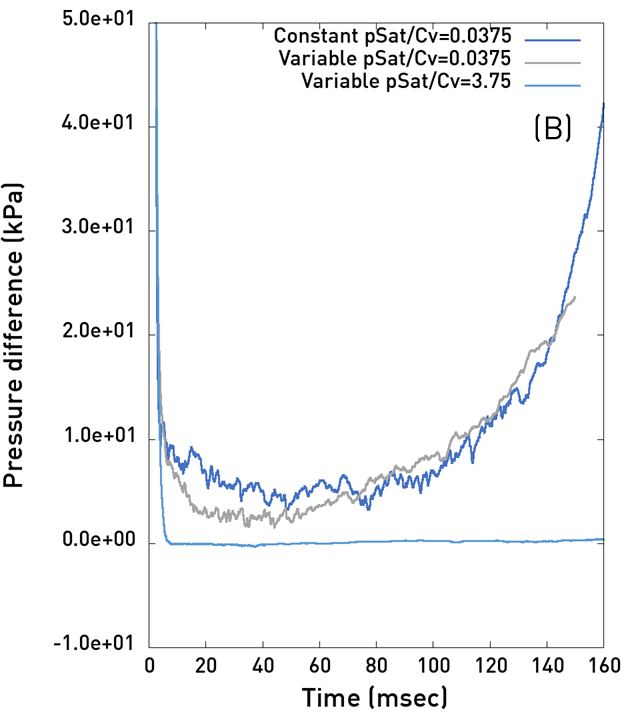}
\end{subfigure}
\caption{Effect of variable saturation pressure on metering and expansion chamber pressure (left) and $\Delta p_{sat}$ pressure differential in the expansion chamber.}
\label{fig:results:temperatureDependentProperties:expansionChamberPressureProfiles}
\end{figure}

\begin{figure}
\centering
\begin{subfigure}{0.495\textwidth}
\centering
\includegraphics[width=\textwidth,trim={0 0 0 0},clip]{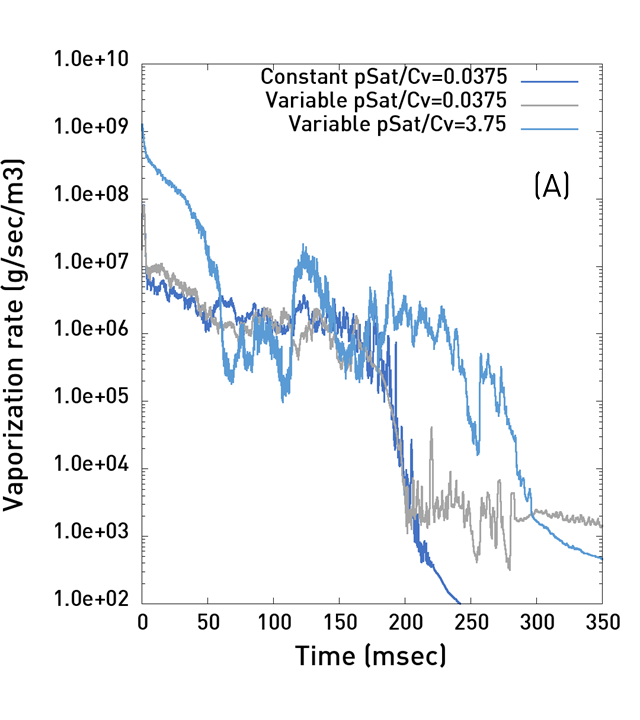}
\end{subfigure}
\hfill
\begin{subfigure}{0.495\textwidth}
\centering
\includegraphics[width=\textwidth,trim={0 0 0 0},clip]{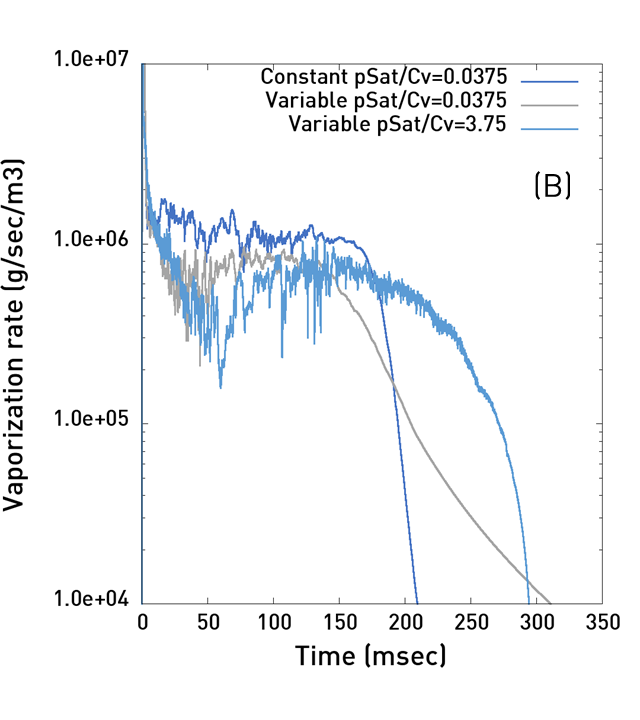}
\end{subfigure}
\caption{Comparison of volumetric vaporization rate in the (a) valve orifice and (b) expansion chamber for constant and variable saturation pressure calculations.}
\label{fig:results:temperatureDependentProperties:componentsVaporizationRates}
\end{figure}

In Fig.~\ref{fig:results:temperatureDependentProperties:expansionChamberPressureProfiles}(a), the calculated mixture pressure inside the metering and expansion chamber is compared to the measurements of Clark~\cite{Clark1991}. In the experiment, a \SI{100}{\micro\liter} metering valve and a valve and nozzle orifice both of 0.26 mm in diameter were used to measure the discharge of a CFC-12 propellant. Therefore, experimental profiles cannot be strictly related to present simulation setup. Nonetheless, they provide a reference for the actual behavior of the mixture pressure in the two chambers. As it can be seen from Fig.~\ref{fig:results:temperatureDependentProperties:expansionChamberPressureProfiles}(a), in the case of constant saturation pressure the mixture pressure in the metering and expansion chamber drops and rises, respectively, almost instantaneously and remains almost constant during the most active part of the flash-boiling ($0\le t\le 150$ \SI{}{\ms}). A significant pressure drop occurs only towards the end of the discharge, when almost all the propellant is vaporized. Using a variable saturation pressure, the initial behavior of the simulated discharge is found in better agreement with the experimental profiles, showing a finite time needed by the metering and expansion chamber pressure to reach the equilibrium value dictated by the saturation pressure, followed by a monotone pressure drop during the remaining of the shot. The analysis of the pressure differential $\left(\Delta p_{sat}=p_{sat}-p\right)$ driving the flash-boiling in the expansion chamber in Fig.~\ref{fig:results:temperatureDependentProperties:expansionChamberPressureProfiles}(b) reveals a reduction in the initial undershoot of pressure below the saturation value in the calculation with variable $p_{sat}$, thereby reducing the amount of vapor production and leading to the increase of mixture density shown in the panel (b) of Fig.~\ref{fig:results:temperatureDependentProperties:nozzleOrificeProfiles}. This is confirmed by the profiles of vaporization rate shown in Fig.~\ref{fig:results:temperatureDependentProperties:componentsVaporizationRates}(b), where a reduction can be seen in the initial stage of the shot ($t<100$ \SI{}{\ms}) with variable saturation pressure.

\begin{figure}
\centering
\includegraphics[width=1.0\textwidth,trim={0 0 0 0},clip]{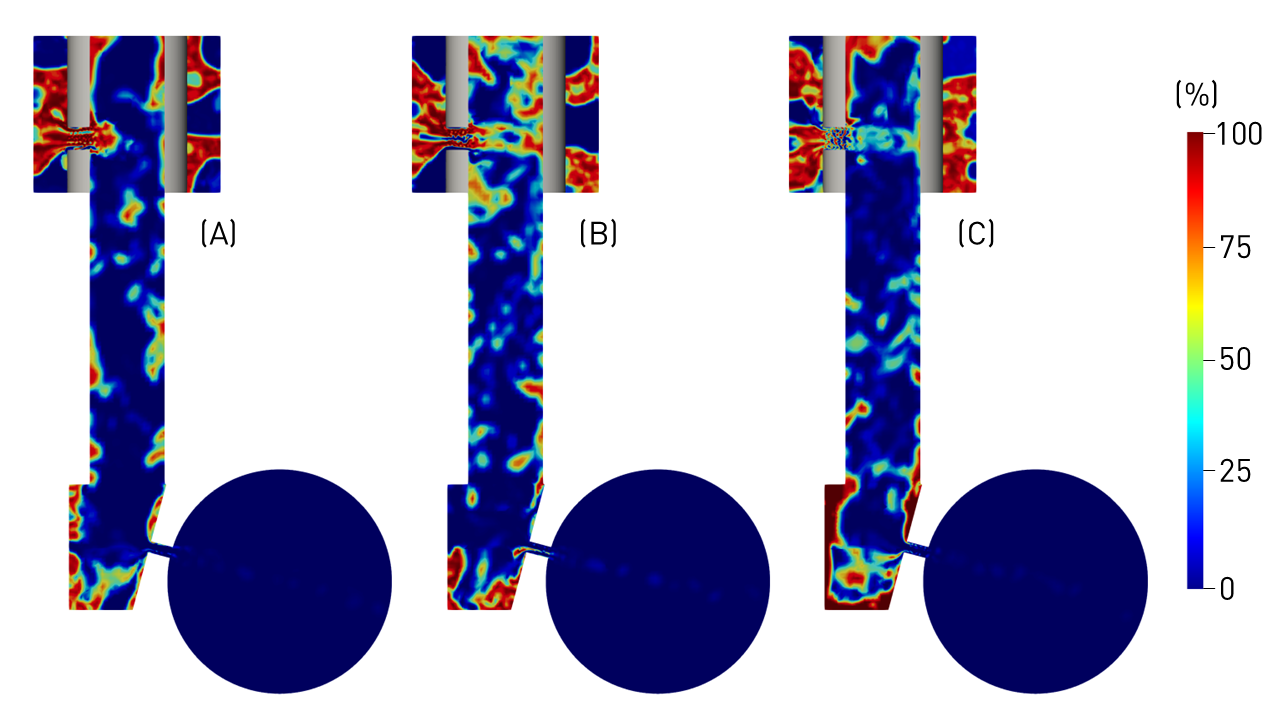}
\caption{Comparison of liquid volume fraction field at $t=30$ \SI{}{\ms}: (a) constant saturation pressure with $C_v=0.0375$, (b) variable saturation pressure with $C_v=0.0375$, (c) variable saturation pressure with $C_v=3.75$.}
\label{fig:results:temperatureDependentProperties:liquidFractionMapComparison}
\end{figure}

An attempt to re-calibrate the flash-boiling model has been performed by increasing the evaporation coefficient to $C_v=3.75$, aiming at compensating for the reduction in vapor production associated with the variable saturation pressure. The results for the mixture density and phases flow rates, also presented in Fig.~\ref{fig:results:temperatureDependentProperties:nozzleOrificeProfiles}, show a significant reduction in the overestimated peak both in the mixture density and liquid flow rate profiles. Nonetheless, the vapor flow rate is still found lower than in the calculation performed with constant saturation pressure and underestimated compared to the measurements of Mason-Smith et al.~\cite{MasonSmith2017}. The analysis of vapor production in Fig.~\ref{fig:results:temperatureDependentProperties:componentsVaporizationRates} reveals a significant increase in the valve orifice for most of the shot compared to the simulation with the standard evaporation coefficient, but similar values in the expansion chamber at the early stage of the simulated metered discharge ($t<50$ \SI{}{\ms}), and even lower values at in the intermediate stage ($50 \le t\le 150$ \SI{}{\ms}). Despite the realistic pressure profile obtained with $C_v=3.75$ shown Fig.~\ref{fig:results:temperatureDependentProperties:expansionChamberPressureProfiles}(a), the pressure differential shown in the panel (b) of the same figure is found to collapse due to the very high evaporation coefficient, suggesting a less realistic behavior of the model. Further indication of the limited reliability of the re-calibrated model can be found in the contour maps of liquid fraction shown in Fig.~\ref{fig:results:temperatureDependentProperties:liquidFractionMapComparison}, where a significant condensation of the propellant, not present in the experimental visualizations, can be noticed in the sump volume.

\subsection{Impact of turbulence modeling}\label{subsection:results:turbulenceModeling}
The impact of the potential onset of a turbulent regime during the metered discharge is investigated in this section by comparing the numerical results obtained with and without the use of a turbulence model. In the turbulent calculations, the widely adopted $k-\omega$ SST model by Menter~\cite{Menter1994} is employed. The simulations are carried out with variable saturation pressure and using the standard evaporation coefficient ($C_v=0.0375$) to explore the effect of turbulent fluctuations on the flash-boiling process using the model of Singhal~\cite{Singhal2002}, presented in Section~\ref{subsection:computationalModel:turbulenceModeling}.

\begin{figure}
\centering
\begin{subfigure}{1.0\textwidth}
\centering
\includegraphics[width=\textwidth,trim={0 0 0 0},clip]{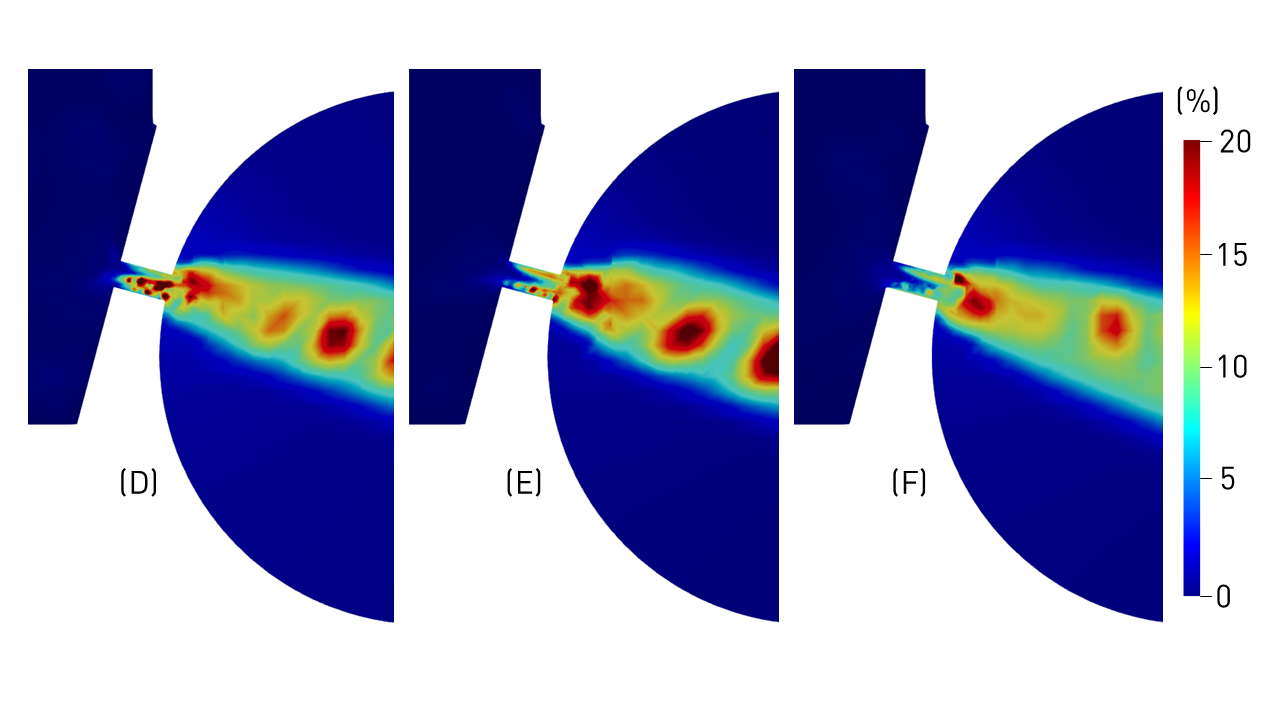}
\end{subfigure}
\hfill
\begin{subfigure}{1.0\textwidth}
\centering
\includegraphics[width=\textwidth,trim={0 0 0 0},clip]{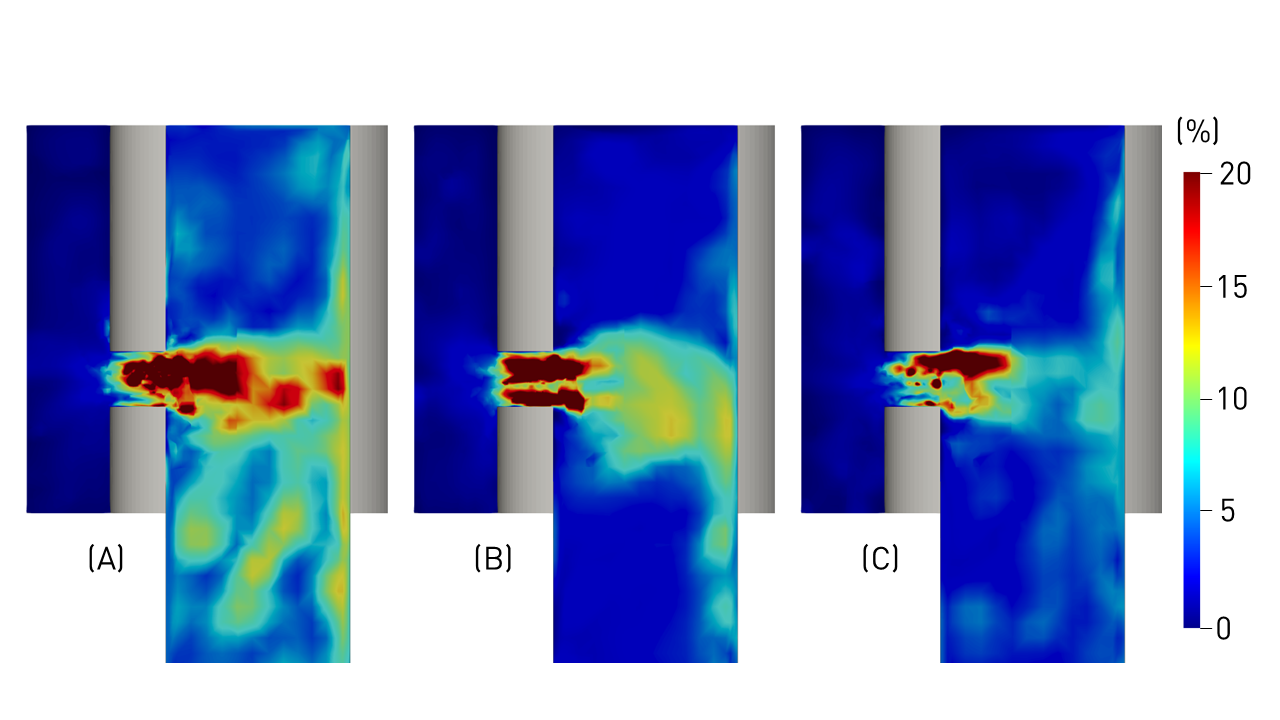}
\end{subfigure}
\caption{Temporal evolution of turbulence intensity in the valve (top) and nozzle (bottom) orifice region: left, $t$=30 \SI{}{\ms}; center, $t$=60 \SI{}{\ms}; right, $t$=90 \SI{}{\ms}.}
\label{fig:results:turbulenceModeling:turbulenceIntensityTimeSequence}
\end{figure}

In Fig.~\ref{fig:results:turbulenceModeling:turbulenceIntensityTimeSequence}, the strength of turbulent velocity fluctuations in the valve orifice and nozzle orifice regions is displayed at selected times during the discharge by plotting the contour maps of turbulence intensity, defined as follows:

\begin{equation}\label{eqn:turbulenceIntensity}
    T_u=\frac{\sqrt{\frac{2}{3}k}}{U}\times100
\end{equation}

where $U$ is the mean velocity characterizing the flow in the corresponding orifice region. High-values of turbulence intensity ($T_u>20\%$) can be seen in the valve orifice volume across the entire active part of the shot ($30\le t\le 90$ \SI{}{\ms}). The low Reynolds number characterizing the flow in the orifice suggests that velocity fluctuations are mainly driven by the flash-boiling taking place inside the orifice (see also Fig.~\ref{fig:results:qualitativeAnalysis:liquidFractionTimeSequenceValveOrifice}(c)), thereby supporting the strong impact of the phase-change process on turbulence production. At the initial stage of discharge ($t=30$ \SI{}{\ms}, see Fig.~\ref{fig:results:turbulenceModeling:turbulenceIntensityTimeSequence}(a)), regions of high-value of turbulence intensity ($T_u>10\%$) are also found in the stem volume, decreasing progressively towards low-levels ($T_u\approx 5\%$) at the beginning of decaying phase of the shot ($t=90$ \SI{}{\ms}, see Fig.~\ref{fig:results:turbulenceModeling:turbulenceIntensityTimeSequence}(c)).

The nozzle-orifice volume is also characterized by very high-values of turbulence intensity in the ramp-up phase of the metered discharge ($t=30$ \SI{}{\ms}, Fig.~\ref{fig:results:turbulenceModeling:turbulenceIntensityTimeSequence}(d)), but velocity fluctuations decreases more rapidly towards low-levels compared to the valve-orifice ($t=30$ \SI{}{\ms}, see Fig.~\ref{fig:results:turbulenceModeling:turbulenceIntensityTimeSequence}(c) and (f)). However, high-values of turbulence intensity are found in the near-orifice region as well as in the region of plume development across the peak of the metered discharge ($30\le t\le 90$ \SI{}{\ms}).

\begin{figure}
\centering
\begin{subfigure}{0.495\textwidth}
\centering
\includegraphics[width=\textwidth,trim={0 0 0 0},clip]{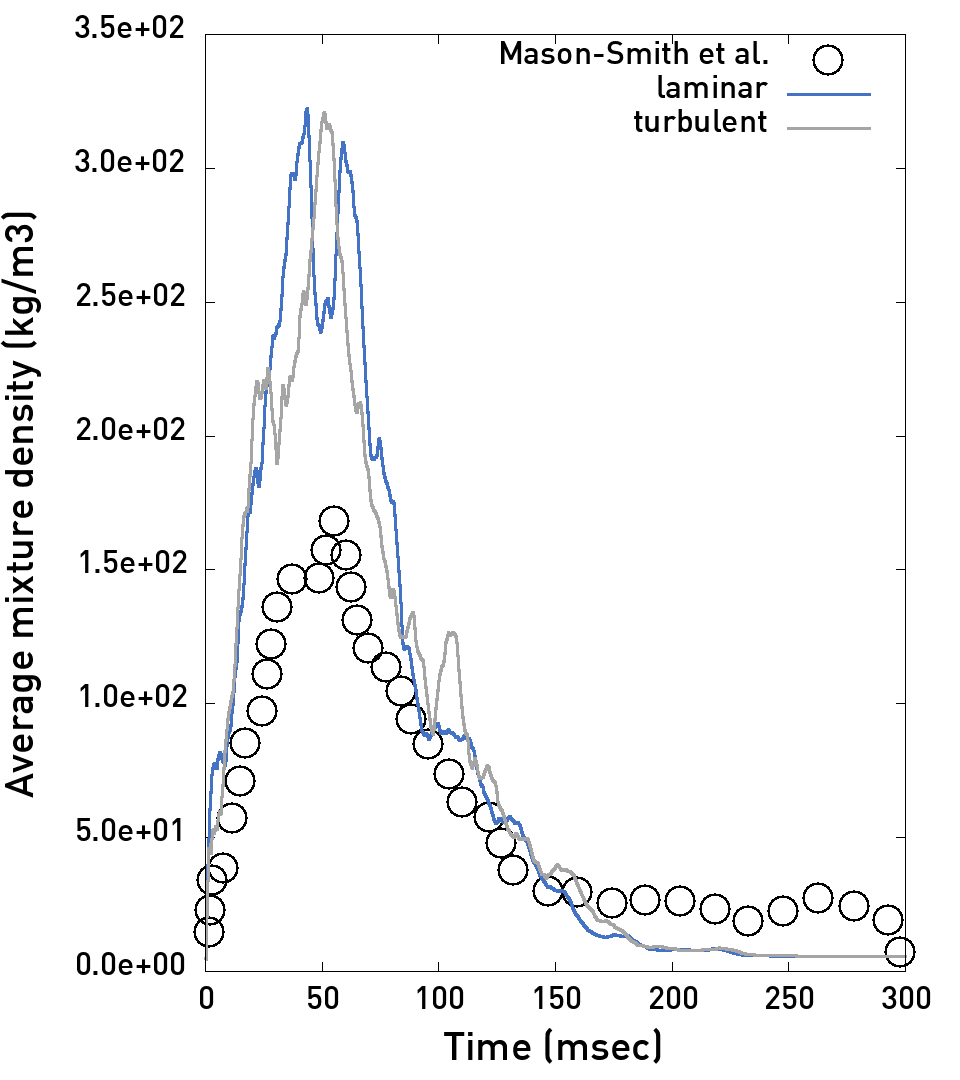}
\end{subfigure}
\hfill
\begin{subfigure}{0.495\textwidth}
\centering
\includegraphics[width=\textwidth,trim={0 0 0 0},clip]{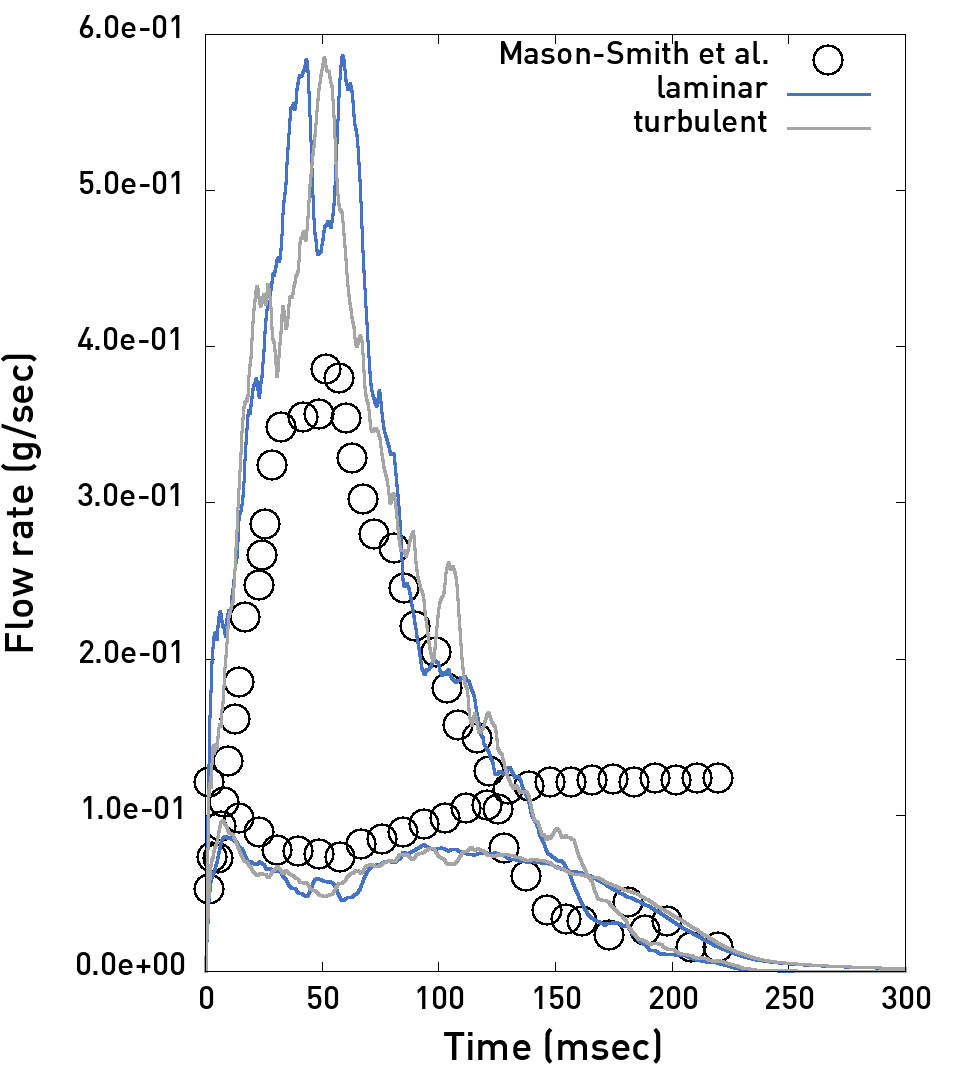}
\end{subfigure}
\caption{Comparison of computed mixture discharge profiles at the nozzle orifice with the measurements of Mason-Smith et al.~\cite{MasonSmith2017} with and without turbulence modeling: (a) mixture density, (b) liquid and vapor flow rates.}
\label{fig:results:turbulenceModeling:nozzleProfilesComparison}
\end{figure}

Despite the high-level of velocity fluctuations characterizing the valve and nozzle orifice regions, particularly at the early stage of the metered discharge, their effect on the predicted profiles of mixture density and liquid and vapor flow rates presented in Fig.~\ref{fig:results:turbulenceModeling:nozzleProfilesComparison} is practically negligible. The most notable difference is limited to a single-peaked vs double-peaked profiles in the turbulent and laminar calculations, respectively. The limited impact of turbulent fluctuations on the metered discharge can partly be explained by the magnitude of pressure fluctuations given by Eq.~(\ref{eqn:pressureFluctuationsModeling}), which is found two orders of magnitudes lower than saturation pressure, even when considering the maximum values of the observed turbulent kinetic energy.

\begin{figure}
\centering
\begin{subfigure}{0.495\textwidth}
\centering
\includegraphics[width=\textwidth,trim={0 0 0 0},clip]{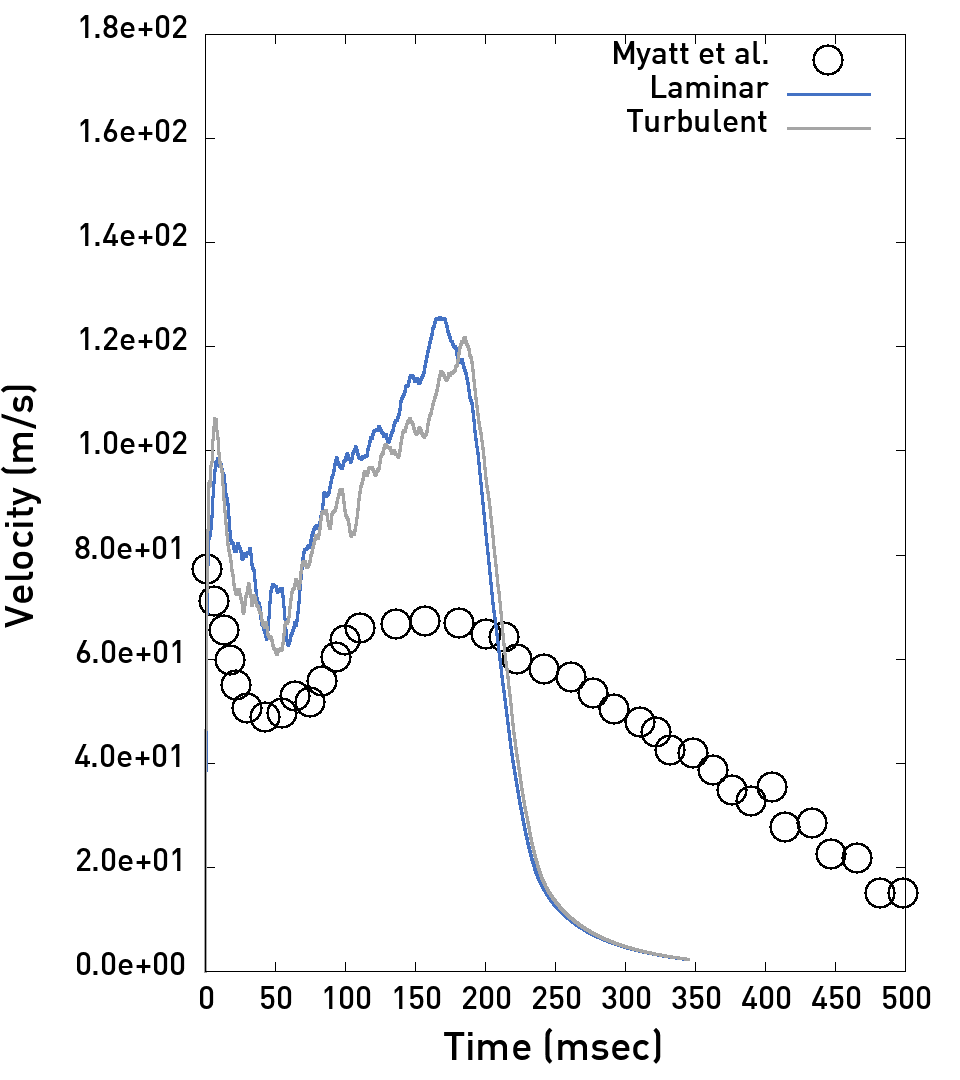}
\end{subfigure}
\hfill
\begin{subfigure}{0.495\textwidth}
\centering
\includegraphics[width=\textwidth,trim={0 0 0 0},clip]{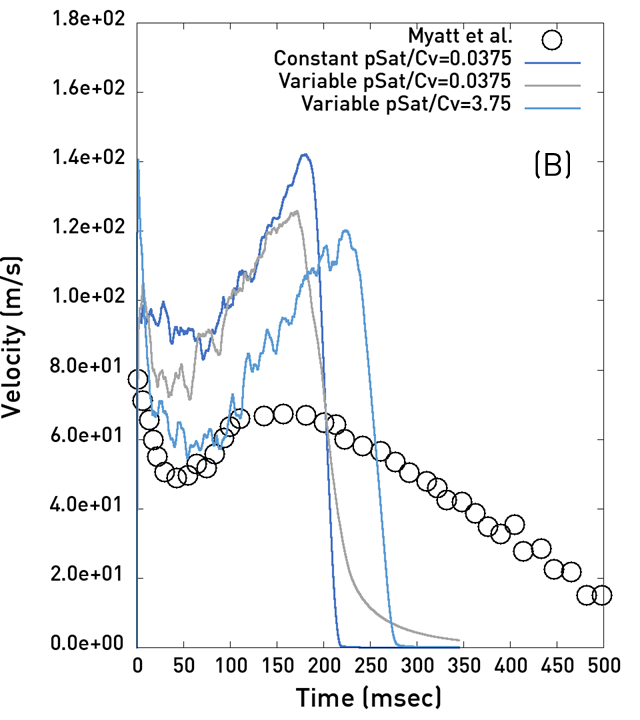}
\end{subfigure}
\caption{Comparison of predicted nozzle-orifice velocity with measurements of Myatt et al.~\cite{Myatt2015a,Myatt2015b} extrapolated in the near-orifice region by~\cite{Gavtash2016}: (a) laminar and turbulent calculations, (b) calculations with constant and variable saturation pressure.}
\label{fig:results:turbulenceModeling:nearNozzleOrificeVelocityComparison}
\end{figure}

Larger differences between the laminar and turbulent calculations can be seen in the nozzle orifice velocity presented in Fig.~\ref{fig:results:turbulenceModeling:nearNozzleOrificeVelocityComparison}(a). Here, a reduction in the orifice velocity up to 20 m/s is observed after the onset of the decaying phase of the metered discharge ($t>50$ \SI{}{\ms}). In the same figure, the measurements performed using Phase-Doppler-Anemometry (PDA) at 15 mm from the nozze orifice by Myatt et al.~\cite{Myatt2015a,Myatt2015b} and extrapolated to the near-orifice region in the work of Gavtash~\cite{Gavtash2016} are also reported. Both the laminar and turbulent calculations overpredict the near-orifice velocity compared to measurements. Nonetheless, the shape of numerical profiles presents a strong similarity to the experiments. The comparison with the predicted nozzle-orifice velocity using a constant saturation pressure in Fig.~\ref{fig:results:turbulenceModeling:nearNozzleOrificeVelocityComparison}(b) shows the initial change in the saturation pressure being essential to capture the local minima at the peak of the shot ($t=50$ \SI{}{\ms}). Moreover, in the same figure the predicted profile with variable saturation pressure and the modified evaporation coefficient ($C_v=3.75$) shows a much better agreement with measurements up to $t=100$ \SI{}{\ms}. The predicted velocity in the remaining of the shot is still overpredicted by the model, but the experiments also show a much longer duration of the shot despite the almost identical device geometry adopted in both the present study and that of Mason-Smith et al.~\cite{MasonSmith2017}. Similar agreement with the experiments and shots durations comparable to the present results are also reported in the work of~\cite{Gavtash2019} using system models, where further evaporation of the mixture due to ambient to actuator heat transfer is indicated as the potential mechanism driving the slower decay of orifice velocity in the experiments. 

\subsection{Mixture temperature predictions}\label{subsection:results:mixtureTemperature}
In the experimental and numerical studies of the pMDI reported in the literature, the main focus is ultimately directed towards the process of aerosol formation and the prediction of droplets size distribution. Equation~(\ref{eqn:ClarkCorrelation}) shows that mixture pressure is the main thermodynamic variable impacting the MMAD of the aerosol and therefore minor attention has generally been paid to the measurement and prediction of the mixture temperature inside the device. The most comprehensive set of measurements available to date is from the work by Clark~\cite{Clark1991}, in which the maximum temperature drop across the valve orifice was measured for various ratios of the nozzle to valve orifice diameters, volumes of the metered chamber and propellants. However, only a single temperature profile from the expansion chamber was reported over time for a CFC-12 propellant and a 100 \SI{}{\micro\liter} metering chamber during the metered discharge. A temperature profile measured during the shot in the sump region was more recently reported by Gavtash et al.~\cite{Gavtash2019} for the same pMDI geometry adopted in the present work, but for a mixture of HFA-134a propellant and 10$\%$ ethanol. A temperature profile for pure HFA-134a propellant was measured in the work of Myatt et al.~\cite{Myatt2022} for the same setup, but only in the nozzle orifice. A comprehensive and consistent set of temperature measurements for a single device geometry and operating conditions is thus missing.

Despite the limited measurements reported in the literature, the temperature inside the pMDI can impact the quality of the drug product. For example, the solubility of the active ingredients dissolved in the mixture are known to be highly sensitive to the mixture temperature, whose variation might lead to earlier precipitation of the ingredients. In this section, the computed temperature profiles are thus compared to available data to further validate the numerical model and assess the predictive capabilities of the simulations for the temperature field.

\begin{figure}
\centering
\begin{subfigure}{0.495\textwidth}
\centering
\includegraphics[width=\textwidth,trim={0 0 0 0},clip]{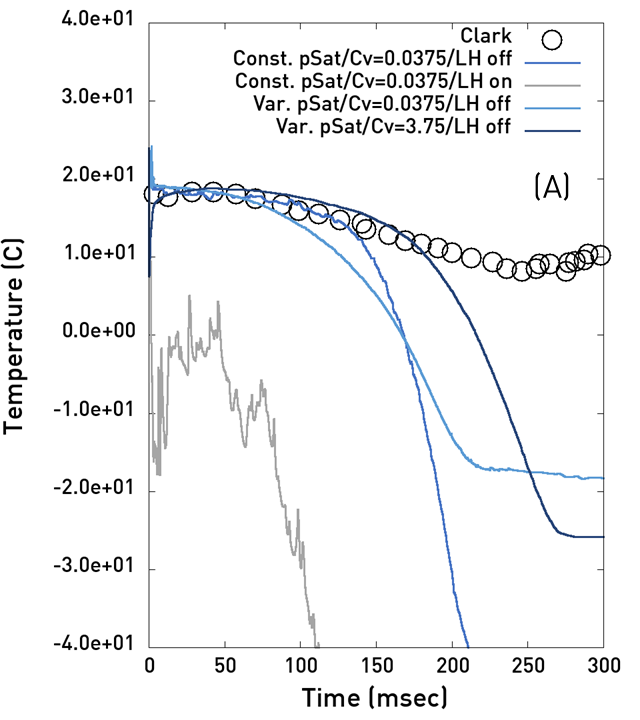}
\end{subfigure}
\hfill
\begin{subfigure}{0.495\textwidth}
\centering
\includegraphics[width=\textwidth,trim={0 0 0 0},clip]{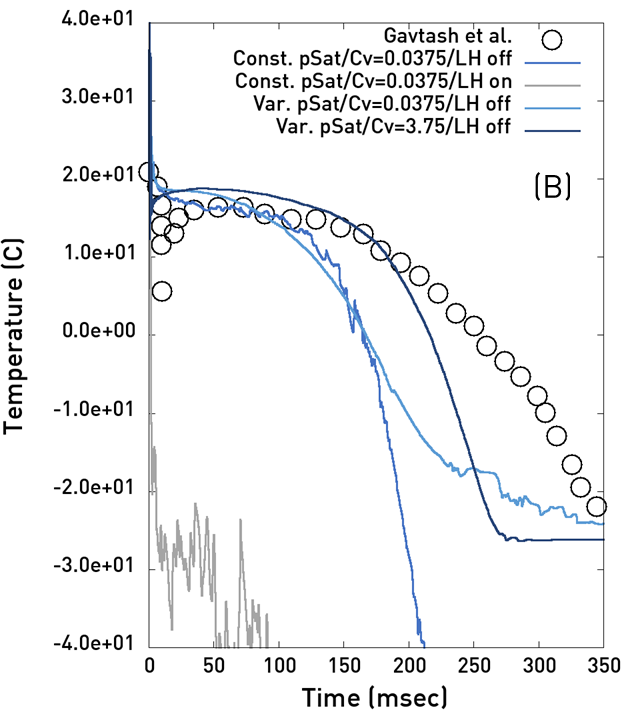}
\end{subfigure}
\begin{subfigure}{0.495\textwidth}
\centering
\includegraphics[width=\textwidth,trim={0 0 0 0},clip]{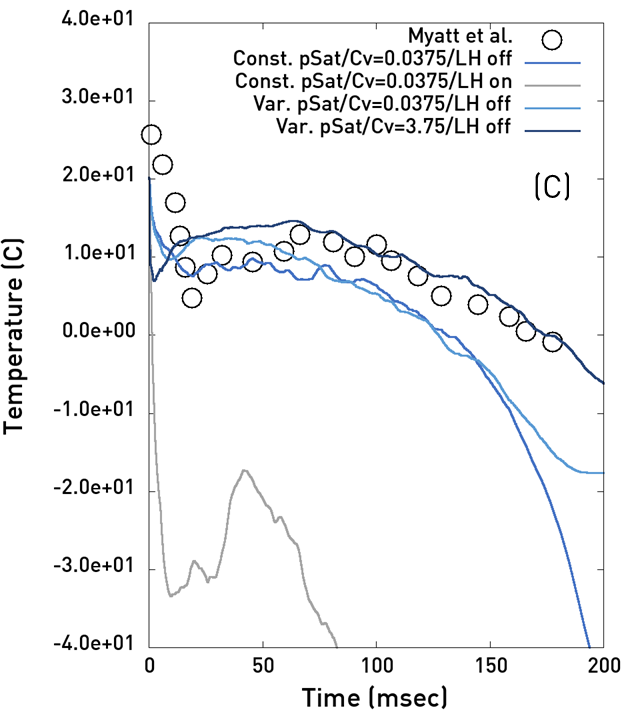}
\end{subfigure}
\caption{Comparison of computed mixture temperature in the (a) stem region, (b) sump (region) and (c) nozzle region with measured data.}
\label{fig:results:temperaturePredictions:mixtureTemperatureComparison}
\end{figure}

In Fig.~\ref{fig:results:temperaturePredictions:mixtureTemperatureComparison}, the predicted mixture temperature profile in the stem and sump regions using various setup of the numerical model are compared to the measurement of Clark~\cite{Clark1991} and Gavtash et al.~\cite{Gavtash2019}. In the calculations, a point probe was defined to record the local mixture temperature during the simulation and mimic the thermocouples mounted inside the device in the experiments. In the stem region, the measured temperature values in the peak region of the shot ($0<t<150$ \SI{}{\ms}) are comparable to those predicted using the constant saturation pressure model with the standard evaporation coefficient ($C_v=0.0375$) and the variable saturation pressure model with the corrected evaporation coefficient ($C_v=3.75$), but without the modeling of the contribution of latent heat to the energy equation. In the simulation with variable saturation pressure, a significant temperature drop of about \SI{10}{\celsius}, which is not observed in the simulation with constant saturation pressure, takes place at the very beginning of the shot, followed by temperature recovery before the monotonic cooling of the mixture occurs during the remaining of the discharge. The same dynamics of the mixture temperature was reported in the work by Myatt~\cite{Myatt2022} for the measurements performed in the nozzle, and explained by the initial expansion of the propellant and flow of cold vapor through the orifice followed by the flow of warmed liquid.

A significant and sudden drop of mixture temperature of the order of \SI{30}{\celsius} downstream of the valve orifice was also reported  by the simulation performed by Clark~\cite{Clark1991} with a system model. However, such a drop cannot be seen in the measurements shown in Fig.~\ref{fig:results:temperaturePredictions:mixtureTemperatureComparison}, suggesting that the sampling rate of the thermocouple adopted in the experiments was too low to capture it. In the present calculations, a temperature drop of similar order of magnitude occurring at the same time scale as in the simulations of Clark~\cite{Clark1991} was obtained by activating the latent heat modeling in the temperature equation. However, no temperature recovery is observed in the simulations and the subsequent monotonic cooling of the mixture, exacerbated by the larger pressure difference between the expansion chamber and the atmosphere associated with the constant saturation pressure (see Fig.~\ref{fig:results:temperatureDependentProperties:expansionChamberPressureProfiles}(a)), leads to unrealistic values of the mixture temperature later in the shot. The effect of the pressure difference is also shown by the temperature profile obtained with constant saturation pressure and without the modeling of latent heat in Fig.~\ref{fig:results:temperaturePredictions:mixtureTemperatureComparison}, where a mixture temperature of \SI{-40}{\celsius} is predicted at $t=200$ \SI{}{\ms}.

A temperature drop of about \SI{15}{\celsius} around $t=10$ \SI{}{\ms} in the sump region can be observed in Fig.~\ref{fig:results:temperaturePredictions:mixtureTemperatureComparison}(b) from the measurements of Gavtash~\cite{Gavtash2019}. Similarly to the temperature profile predicted by the present model in the stem region, only the simulation with variable saturation pressure and the corrected evaporation coefficient is able to represent the temperature drop which, however, takes place almost immediately after the beginning of the shot and thus much earlier than in the experiments. The subsequent temperature recovery and monotonic drop up to $t=200$ \SI{}{\ms} as well as the asymptotic mixture temperature at the end of the shot are in fair agreement with measurements.

A comparison of the present simulations with the measurements performed in the nozzle region by Myatt~\cite{Myatt2022} is presented in Fig.~\ref{fig:results:temperaturePredictions:mixtureTemperatureComparison}(c). In this case, a moving-average filter is applied to the calculated temperature profiles due to significant oscillations occurring in the flow inside the nozzle orifice. The experiments, performed in very similar conditions to those reported by Gavtash~\cite{Gavtash2019}, present an initial temperature drop of about \SI{20}{\celsius} taking place around $t=20$ \SI{}{\ms}. While the simulations with constant and variable saturation pressure with the standard and the corrected evaporation coefficient both predict a similar temperature drop, which is found underpredicted compared to the experiments, only the calculation with variable saturation pressure gives a temperature recovery similar to measurements and also a good agreement for the monotonic decay of mixture temperature.

\section{Impact of geometry modifications}
One of the main advantages of CFD models over system models is the ability to explicitly capture the effect of geometrical modifications, thereby allowing for a better evaluation of the effect of such changes on device actuation and performance. An example is given by the work of Duke et al.~\cite{Duke2021}, where a CFD model was developed and used in combination with experimental techniques to explore novel actuator designs. The model, also based on the open-source CFD software OpenFOAM and adopting the HRM model to represent the phase-change, was used to show the effect of replacing a single nozzle orifice with a twin-orifice system on the aerosol formation and on characteristics of the spray emitted by the pMDI. Given the focus on the external flow downstream of the nozzle orifice, the metered discharge was not represented explicitly in the analysis and the metering valve simulated with a time-varying pressure inlet.

\begin{figure}
\centering
\includegraphics[width=1.0\textwidth,trim={0 0 0 0},clip]{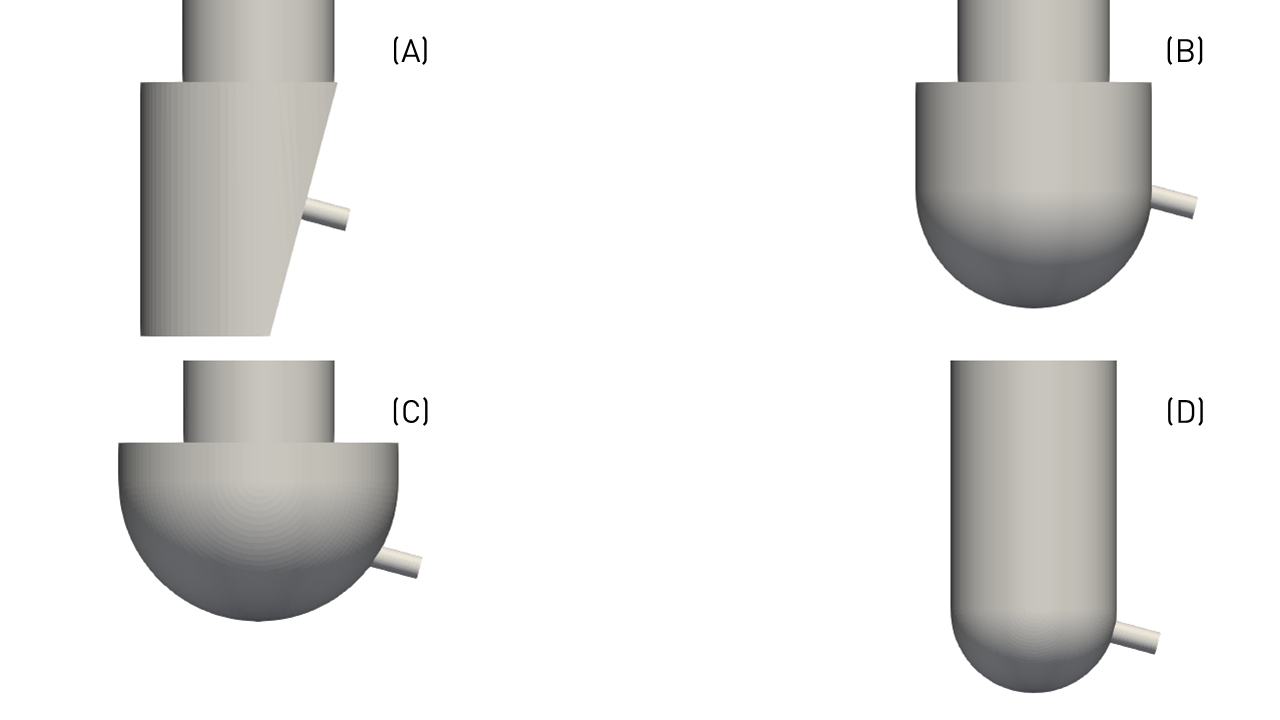}
\caption{Comparison of standard (A/STD) and modified (B/SUMP01, C/SUMP02, D/SUMP03) sump geometries.}
\label{fig:results:geometryModifications:sumpGeometries}
\end{figure}

Here, we present a comparison of different sump designs to explore their effect on mixture profiles in the nozzle orifice as well as on the mixture flow in the sump volume. The modified sump geometries, shown in Fig.~\ref{fig:results:geometryModifications:sumpGeometries}, are characterized by a rounded shape of different diameters and depths (see Tab.~\ref{tab:results:geometryModifications:sumpDimensions}) but share the same volume of $18.8$ \SI{}{\micro\liter} characterizing the standard design. Moreover, while the SUMP01 and SUMP02 also share a sudden expansion between the stem and sump volumes with the standard geometry, the SUMP03 design has a uniform diameter and much higher depth. The simulations are performed using the medium size grid with boundary layers, also adopted in the turbulent calculations, to improve the resolution in the near-wall region of the sump. The standard evaporation coefficient ($C_v=0.0375$) and a constant saturation pressure are employed in the analysis and the latent heat source is not taken into account in the energy equation.

\begin{table}
\begin{center}
\begin{tabular}{ccc}
Sump geometry & Diameter (mm) & Depth (mm) \\ 
\hline
STD    & 3.12 & 3.35 \\
SUMP01 & 3.12 & 3 \\ 
SUMP02 & 3.7  & 2.4 \\  
SUMP03 & 2.2  & 5.3
\end{tabular}
\caption{Characteristic dimensions of modified sump geometries.}
\label{tab:results:geometryModifications:sumpDimensions}
\end{center}
\end{table}

\begin{figure}
\centering
\begin{subfigure}{0.495\textwidth}
\centering
\includegraphics[width=\textwidth,trim={0 0 0 0},clip]{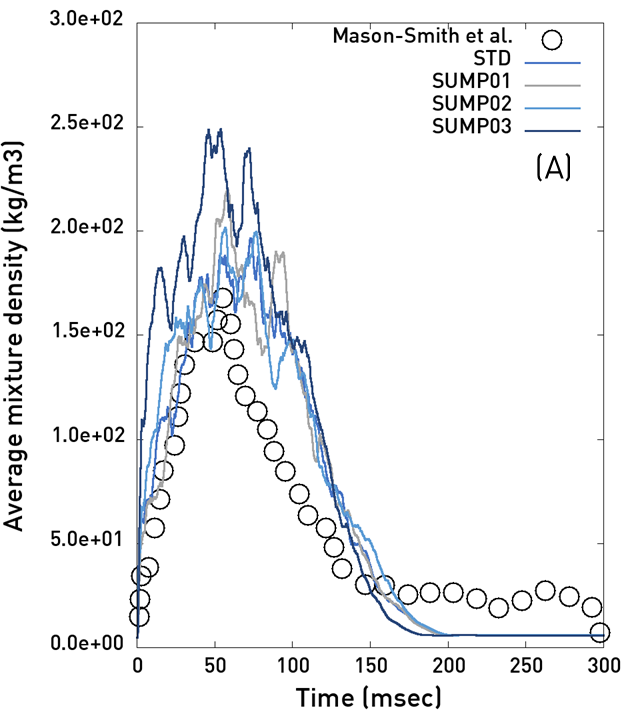}
\end{subfigure}
\hfill
\begin{subfigure}{0.495\textwidth}
\centering
\includegraphics[width=\textwidth,trim={0 0 0 0},clip]{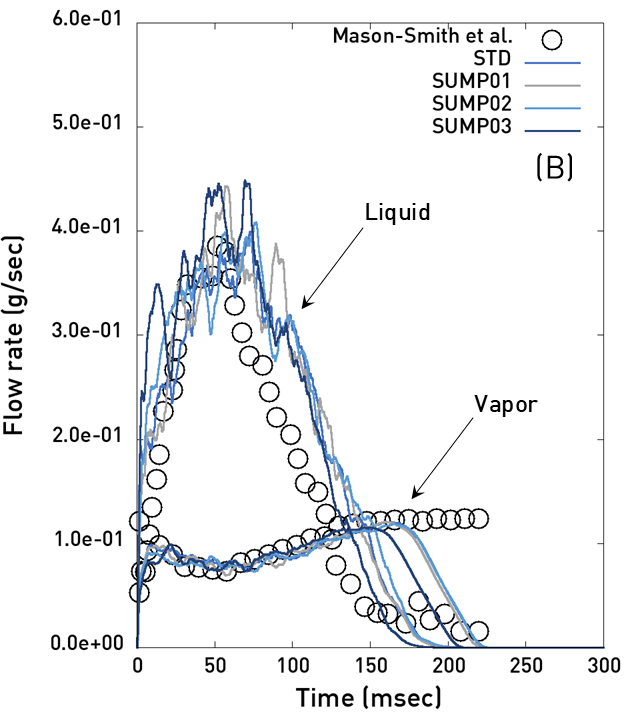}
\end{subfigure}
\caption{Comparison of computed mixture discharge profiles at the nozzle orifice with the measurements of Mason-Smith et al.~\cite{MasonSmith2017} for different sump geometries: (a) mixture density, (b) liquid and vapor flow rates.}
\label{fig:results:geometryModifications:nozzleProfilesComparison}
\end{figure}

\begin{figure}
\centering
\includegraphics[width=1.0\textwidth,trim={0 0 0 0},clip]{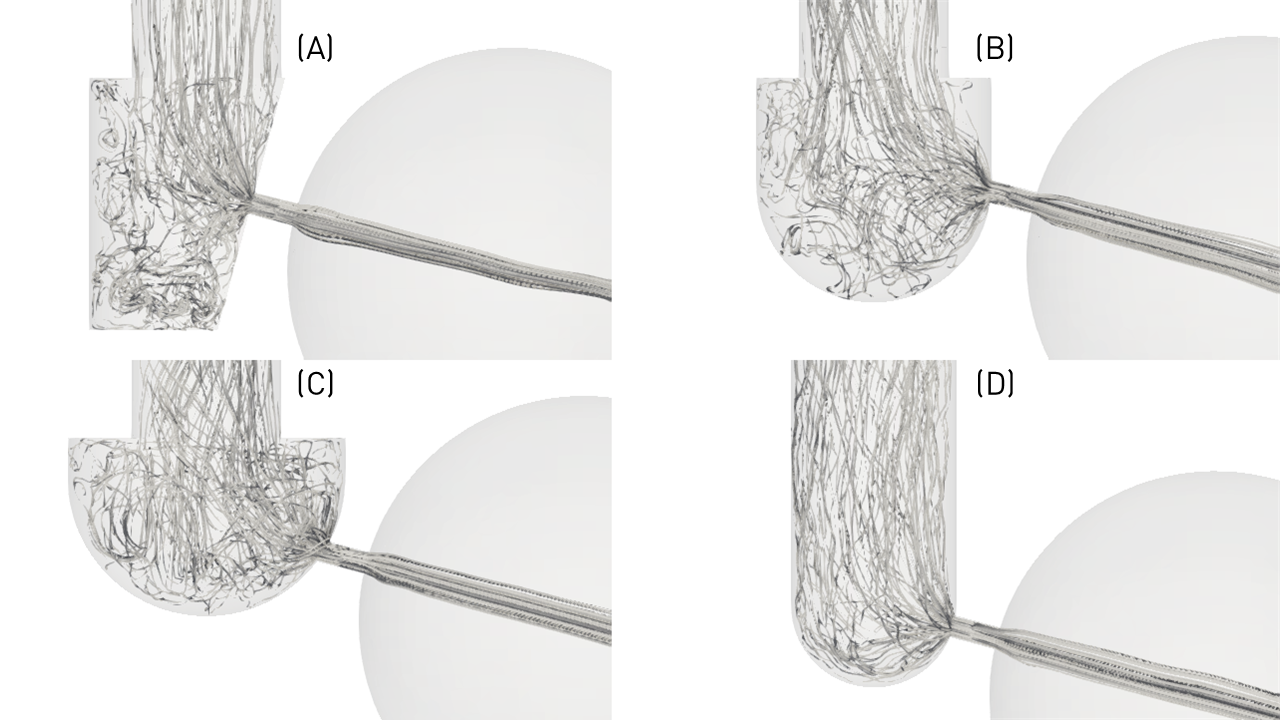}
\caption{Comparison of mixture flow structure in the sump volume at $t=25$ \SI{}{\ms}: (a) standard sump geometry, (b) SUMP01, (c) SUMP02, (d) SUMP03.}
\label{fig:results:geometryModifications:sumpStreamlinesComparison}
\end{figure}

Mixture profiles at the nozzle orifice associated with the metered discharge through the standard and modified sump geometries are compared in Fig.~\ref{fig:results:geometryModifications:nozzleProfilesComparison}. Similar profiles are found between the standard and SUMP01/SUMP02 modified geometries for the mixture density in panel (a), whereas the SUMP03 presents a much more rapid increase in mixture density as well a reduction in the duration of the shot of about 15 \SI{}{\ms} thanks to the very smooth flow taking place in the sump volume (see Fig.~\ref{fig:results:geometryModifications:sumpStreamlinesComparison}). The SUMP03 design also shows larger oscillations in the profile at lower frequencies compared to the other geometries. Such low frequency oscillations are also seen in the liquid phase flow rate in panel (b) of Fig.~\ref{fig:results:geometryModifications:nozzleProfilesComparison}, and to some extent in that of the vapor phase. Both profiles also highlight the shorter duration of the shot, but no significant differences are found in the phases flow rates between the standard and the modified sump geometries.

Despite the similarity between the nozzle-orifice mixture profiles, the modified designs lead to major changes in the flow structure inside the sump volume as shown by Fig.~\ref{fig:results:geometryModifications:sumpStreamlinesComparison}, where the streamlines characterizing the mixture flow are compared at $t=25$ \SI{}{\ms} during the discharge. In the SUMP03 geometry, a much more regular flow is observed, filling almost entirely the sump volume and preventing the onset of major flow recirculations, whereas the standard geometry presents a complex and irregular flow with  vortical structures of several length-scales. Large flow recirculations are also observed in the SUMP02 and SUMP03 geometries, but the flow presents a more regular structure overall.

\begin{figure}
\centering
\begin{subfigure}{0.9\textwidth}
\centering
\includegraphics[width=\textwidth,trim={50 0 50 0},clip]{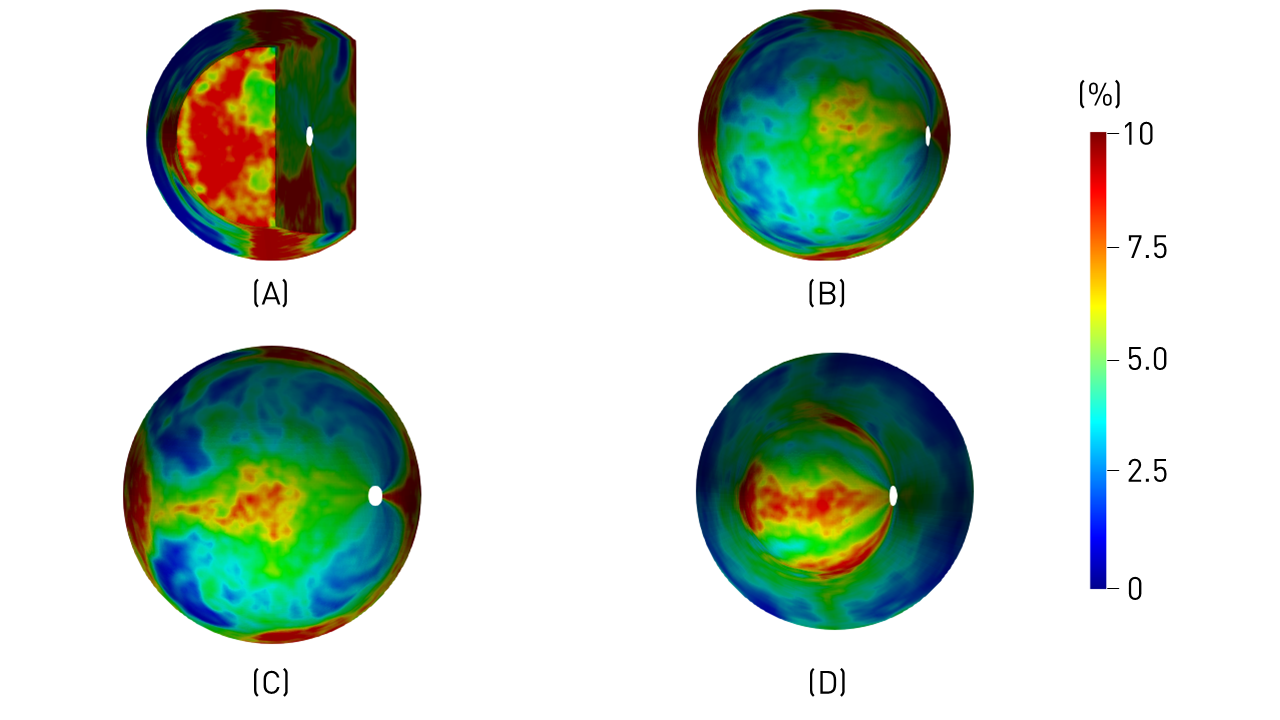}
\end{subfigure}
\hfill
\begin{subfigure}{0.9\textwidth}
\centering
\includegraphics[width=\textwidth,trim={50 0 50 0},clip]{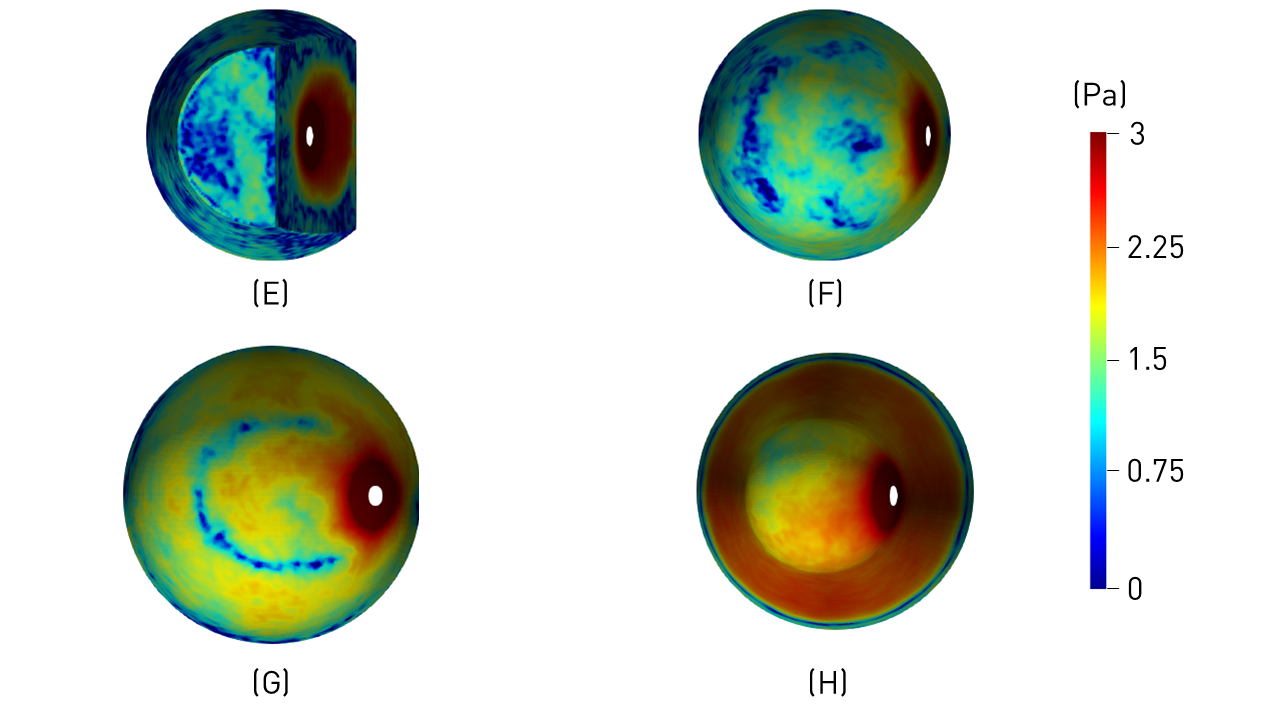}
\end{subfigure}
\caption{Comparison of average liquid fraction (A-D) and average wall shear-stress (E-H) on the sump surface during the metered discharge: (A,E) STD, (B,F) SUMP01, (C,G) SUMP02, (D,H) SUMP03.}
\label{fig:results:geometryModifications:sumpSurfaceStatisticsComparison}
\end{figure}

The different flow structure associated with the standard and modified sump geometries can have a major impact on stagnation of the liquid phase in the sump volume at the end of shot which, along with other factors, is known to potentially lead to the clogging phenomenon of the device due to accumulation of the active ingredient upon evaporation of the liquid phase. In Fig.~\ref{fig:results:geometryModifications:sumpSurfaceStatisticsComparison}, the average liquid fraction on the sump surface in the last 100 \SI{}{\ms} of the metered discharge is shown for the different sump geometries. In the case of the SUMP03 geometry, the shorter duration of the shot is taken into account and the average liquid fraction computed over a temporal window of the same size but shifted 15 \SI{}{\ms} earlier compared to the other geometries. The contour maps of the average liquid fraction shows a significantly higher liquid accumulation in the case of the standard geometry. The comparison with the modified sump geometries SUMP01/SUMP02 reveals how the hard edges of the standard sump are responsible for the liquid accumulation. This is also shown by the contour map of the SUMP03 geometry, where the significantly smaller curvature of the sump compared to SUMP01 and SUMP02 leads to an increase of the average local liquid fraction.

The increased liquid accumulation can be partly associated with the average shear-stress induced by the flow on the sump surface during the shot, also shown in Fig.~\ref{fig:results:geometryModifications:sumpSurfaceStatisticsComparison}. where the local high-values of average liquid fraction in the standard geometry correspond to the local reduction in the average shear stress. On the other hand, the spherical designs lead to significantly higher values of average wall shear-stress, promoting a self-cleaning effect during the device operation.

\section{Conclusions}\label{conclusions}
The CFD model presented in this work represents a first attempt in the explicit simulation of the flashing-flow inside a pMDI. By employing an existing cavitation model to account for the onset of flash boiling upon device actuation, and the VOF technique to represent the resulting liquid–vapor mixture, the model is shown to provide a reasonably accurate prediction of the metered discharge with a computational effort compatible with industrial applications.

The comparison with X-ray visualizations of the mixture flow reveals strong similarity between the numerical solution and the experiments, with the model, providing further insights into the dynamics of the flow across the valve orifice and in the expansion chamber. The validation of the model, carried out primarily by comparing the predicted profiles of mixture density and phase flow rates through the nozzle orifice with measurements based on quantitative radiography, shows a reasonable agreement with the experiments after tuning the evaporation coefficient of the flash-boiling model. The sensitivity analysis to the main modeling parameters, such as the grid resolution and temperature dependence of mixture properties, shows strong robustness from the model, with no major changes in the overall trends of predicted profiles.

Among the modeling parameters and for a fixed value of the evaporation coefficient, the temperature dependence of the propellant saturation pressure has the largest impact on numerical results. Despite the use of a constant saturation pressure gives better agreement with the measured mixture density and flow rates at the nozzle orifice, particularly for the vapor phase, a more consistent behavior of the model using a variable saturation pressure was observed, overall, when comparing the numerical results to additional measurements for the mixture pressure in the expansion chamber and for the near-orifice mixture velocity downstream of the nozzle. 

Temperature profiles represent the main uncertainty in the simulated metered-discharge due to the limited and inconsistent set of measurements available in the literature, where significant differences can be found between experiments performed for similar device geometries and operating conditions. This also leaves some open questions concerning the modeling of the propellant thermodynamics, where the use of a source term in the energy equation to account for the effect of latent heat during the flash-boiling led to unrealistic temperature drops during the metered discharge, while a fair agreement with temperature measurements was obtained, overall, when accounting for temperature variations due to compressible effects only.

Compared to existing system models, which are able to estimate the mixture properties at the nozzle-orifice with a similar degree of accuracy but at a much lower computational cost, the present work demonstrated how the CFD model can be used to study the effect of design changes by comparing modified sump geometries to the standard design. The analysis highlighted the impact of the sump geometry on the residual liquid fraction on the sump surface at the end of the shot and the potential of the modified design to reduce the chance of clogging without altering, at the same time, the average mixture density and flow rates at the nozzle orifice. The present model, together with the CFD modeling of the spray emitted by the device for which the internal flow simulation can work as a precursor calculation, thus represent a first successful step towards the development of a complete digital-twin of the pMDI.

\section*{Acknowledgments}
The authors wish to thank Prof. F. Piscaglia from the Politecnico di Milano, Italy, for providing some guidelines in the initial development of the model and M. G. De Giorgi from the University of Salento, Italy, for the useful discussion on the implementation of the latent heat source in the energy equation of the custom solver.

\section*{CRediT authorship contribution statement}
R.Rossi: development of numerical model, methodology, investigation, formal analysis, validation, visualizations, writing original draft. A. Benassi: conceptualization, supervision, writing original draft, review \& editing, conceptualization.  C. Cottini: conceptualization, supervision, writing, review \& editing, project administration.

\section*{Declaration of Competing Interest}
A. Benassi and C. Cottini are Chiesi Farmaceutici employees. R.Rossi is a consultant/contractor currently engaged by Chiesi Farmaceutici.

\appendix

\bibliographystyle{elsarticle-num} 
\bibliography{references}





% \end{thebibliography}
\end{document}